\documentclass[twocolumn,dvipsnames]{aastex62}
\usepackage{xcolor}
\usepackage{amsmath}
\usepackage{appendix}
\usepackage{makecell}%To keep spacing of text in tables
\usepackage{relsize}
\setcellgapes{4pt}%parameter for the spacing
\usepackage[some]{background}
\usepackage[caption=false, position=top]{subfig}
\usepackage{hyperref}
\usepackage{xspace}

\SetBgScale{3}
\SetBgContents{{%
  {\Huge PRELIMINARY}}}
\SetBgColor{gray}
\SetBgAngle{-45}
\SetBgOpacity{0.5}

\newcommand{\chieff}{\ensuremath{\chi_{\rm eff}}\xspace}

\newcommand{\FracBelow}{74\%}
\newcommand{\meanObsSlopeBase}{-0.1}
\newcommand{\meanObsSlopeEvol}{-0.46}
\newcommand{\meanMockSlopeEvol}{-0.49}
\newcommand{\noEvolMean}{0.06^{+0.05}_{-0.05}}
\newcommand{\noEvolSigma}{0.12^{+0.05}_{-0.05}}
\newcommand{\noEvolBetaq}{1.1^{+3.0}_{-1.4}}

\newcommand{\percentAlphaNegative}{98.7\%}
\newcommand{\percentAlphaNegativeNoOhFiveSeventeen}{99.1\%}
\newcommand{\percentAlphaNegativeNoOhFourTwelve}{99.0\%}
\newcommand{\percentAlphaNegativeNoOhFourTwelveOrOhFiveSeventeen}{95.2\%}
\newcommand{\percentAlphaNegativeWithOhEightFourteen}{92.6\%}
\newcommand{\percentBetaNegativeWithOhEightFourteen}{96.8\%}
\newcommand{\injMean}{0.05^{+0.09}_{-0.09}}
\newcommand{\injSigma}{0.13^{+0.09}_{-0.07}}
\newcommand{\injQuantileAtOrigin}{38}

\newcommand{\evidenceNoEvolNoNeg}{0.00}
\newcommand{\evidenceNoEvolNoNegError}{0.04}
\newcommand{\evidenceNoEvolYesNeg}{1.29}
\newcommand{\evidenceNoEvolYesNegError}{0.04}
\newcommand{\evidenceYesEvolNoNeg}{3.54}
\newcommand{\evidenceYesEvolNoNegError}{0.03}
\newcommand{\evidenceYesEvolYesNeg}{3.64}
\newcommand{\evidenceYesEvolYesNegError}{0.03}
\newcommand{\bayesNegWhenCorrelated}{0.10}

\newcommand{\oddsCorrelatedVsUncorrelated}{\ensuremath{10.5}\xspace}

\newcommand{\LIGOlabMIT}{\affiliation{LIGO Laboratory, Massachusetts Institute of Technology, 185 Albany St, Cambridge, MA 02139, USA}}
\newcommand{\MKI}{\affiliation{Department of Physics and Kavli Institute for Astrophysics and Space Research, Massachusetts Institute of Technology, 77 Massachusetts Ave, Cambridge, MA 02139, USA}}
\newcommand{\CCA}{\affiliation{Center for Computational Astrophysics, Flatiron Institute, 162 Fifth Avenue, New York, NY 10010, USA}}
\newcommand{\SBU}{\affiliation{Department of Physics and Astronomy, Stony Brook University, Stony Brook NY 11794, USA}}

\shortauthors{Callister et al.}

\begin{document}

\title{Who Ordered That? Unequal-mass binary black hole mergers have larger effective spins}

\correspondingauthor{T. Callister}
\email{tcallister@flatironinstitute.org\\
haster@mit.edu\\
kenkyng@mit.edu\\
salvo@mit.edu\\
will.farr@stonybrook.edu}

\author[0000-0001-9892-177X]{Thomas A. Callister}
\CCA{}

% You can add your ORCiD number to the \author[ORCiD]{name} command like I've done below, and it'll be linked automagically later on :)
\author[0000-0001-8040-9807]{Carl-Johan Haster}
\LIGOlabMIT
\MKI

\author[0000-0003-3896-2259]{Ken K. Y. Ng}
\LIGOlabMIT
\MKI

\author[0000-0003-2700-0767]{Salvatore Vitale}
\LIGOlabMIT
\MKI

\author[0000-0003-1540-8562]{Will M. Farr}
\CCA{}
\SBU{}

%TC:ignore
\begin{abstract}
Hierarchical analysis of the binary black hole (BBH) detections by the Advanced LIGO and Virgo detectors has offered an increasingly clear picture of their mass, spin, and redshift distributions.
Fully understanding the formation and evolution of BBH mergers will require not just the characterization of these marginal distributions, though, but the discovery of any correlations that exist between the properties of BBHs.
Here, we hierarchically analyze the ensemble of BBHs discovered by the LIGO and Virgo with a model that allows for intrinsic correlations between their mass ratios $q$ and effective inspiral spins $\chieff$.
At $\percentAlphaNegative$ credibility, we find that the mean of the $\chieff$ distribution varies as a function of $q$, such that more unequal-mass BBHs exhibit systematically larger $\chieff$.
We find Bayesian odds ratio of \oddsCorrelatedVsUncorrelated in favor of a model that allows for such a correlation over one that does not.
Finally, we use simulated signals to verify that our results are robust against degeneracies in the measurements of $q$ and $\chi_\mathrm{eff}$ for individual events.
While many proposed astrophysical formation channels predict some degree correlation between spins and mass ratio, these predicted correlations typically act in an opposite sense to the trend we observationally identify in the data.
\vspace{1cm}
\end{abstract}
%TC:endignore

%%%%%%%%%%%%%%%%%%%%%%
\section{Introduction}
\label{sec:intro}
%%%%%%%%%%%%%%%%%%%%%%

%\BgThispage

The growing number of gravitational-wave detections made by the Advanced
LIGO~\citep{aligo} and Advanced Virgo~\citep{avirgo} observatories is enabling
exploration of the stellar-mass compact binary population at an ever
accelerating pace. With data now available from the first three LIGO-Virgo
observing runs~\citep{gwtc1,gwtc2}, we are beginning to resolve interesting
features in the mass, spin, and redshift distributions of binary black holes
(BBHs)~\citep{O3a_pop,Roulet2020}. The BBH primary mass spectrum is
characterized by a power law at low masses and a possible ``bump'' near
$40\,M_\odot$, followed by a steeper decline and a possible secondary feature near
$80\,M_\odot$~\citep{Fishbach2017,Wysocki2019,Kimball2020,Roulet2020,Tiwari2020,O3a_pop}.
Black hole spins appear to be small but non-zero, and are oriented neither
isotropically nor strictly parallel to the binaries' orbits but with some spread
in spin-orbit tilt
angles~\citep{Farr2017,Farr2018,Tiwari2018,Roulet2019,Wysocki2019,Biscoveanu2020,Miller2020}. Meanwhile, the
BBH merger rate likely increases with redshift at a rate comparable to cosmic
star formation~\citep{Fishbach2018,Callister2020,O3a_pop,O3_isotropic}.

The distributions of black hole parameters encode a valuable range of
astrophysical information, and may elucidate the processes
governing compact binary birth and evolution. In addition to features in these
one-dimensional distributions, \textit{correlations} between parameters will be
particularly  valuable to identify and understand. Different proposed formation
channels, for instance, predict a variety of distinctive correlations that may
exist among compact binaries. One might, for example, expect a correlation
between BBH mass and spin if black holes experience repeated hierarchical
mergers in dense stellar
environments~\citep{Portegies2002,Mckernan2012,Antonini2016,Fishbach2017_hierarchical,Gerosa2017,Doctor2019,Rodriguez2019,Kimball2020}.
And the strength of tidal interactions among field binaries might conceivably
regulate BH spins in a way that depends on their mass
ratio~\citep{Hotokezaka2017,Gerosa2018,Qin2018,Zaldarriaga2018,Bavera2020,Bavera2021}.

Here, we report an apparent anti-correlation between the mass ratio and
effective inspiral spins of binary black hole mergers: more extreme mass ratios
correspond to larger effective spins. At mass ratios near unity, the effective
spins of BBHs are consistent with a narrow distribution that is symmetric about
zero. At unequal mass ratios, however, the BBH population exhibits
preferentially positive effective spins, due to an overall
shift of the effective spin distribution towards larger values.
This behavior is generally inconsistent with predictions of the standard models of compact object formation discussed above.

%%%%%%%%%%%%%%%%%%%%%%
\section{Residual structure in the mass ratio \& effective spin plane}
\label{sec:residuals}
%%%%%%%%%%%%%%%%%%%%%%

\begin{figure*}[t!]
    \centering
    \subfloat[\label{fig:default-posteriors-a}]{\includegraphics[width=0.48\textwidth]{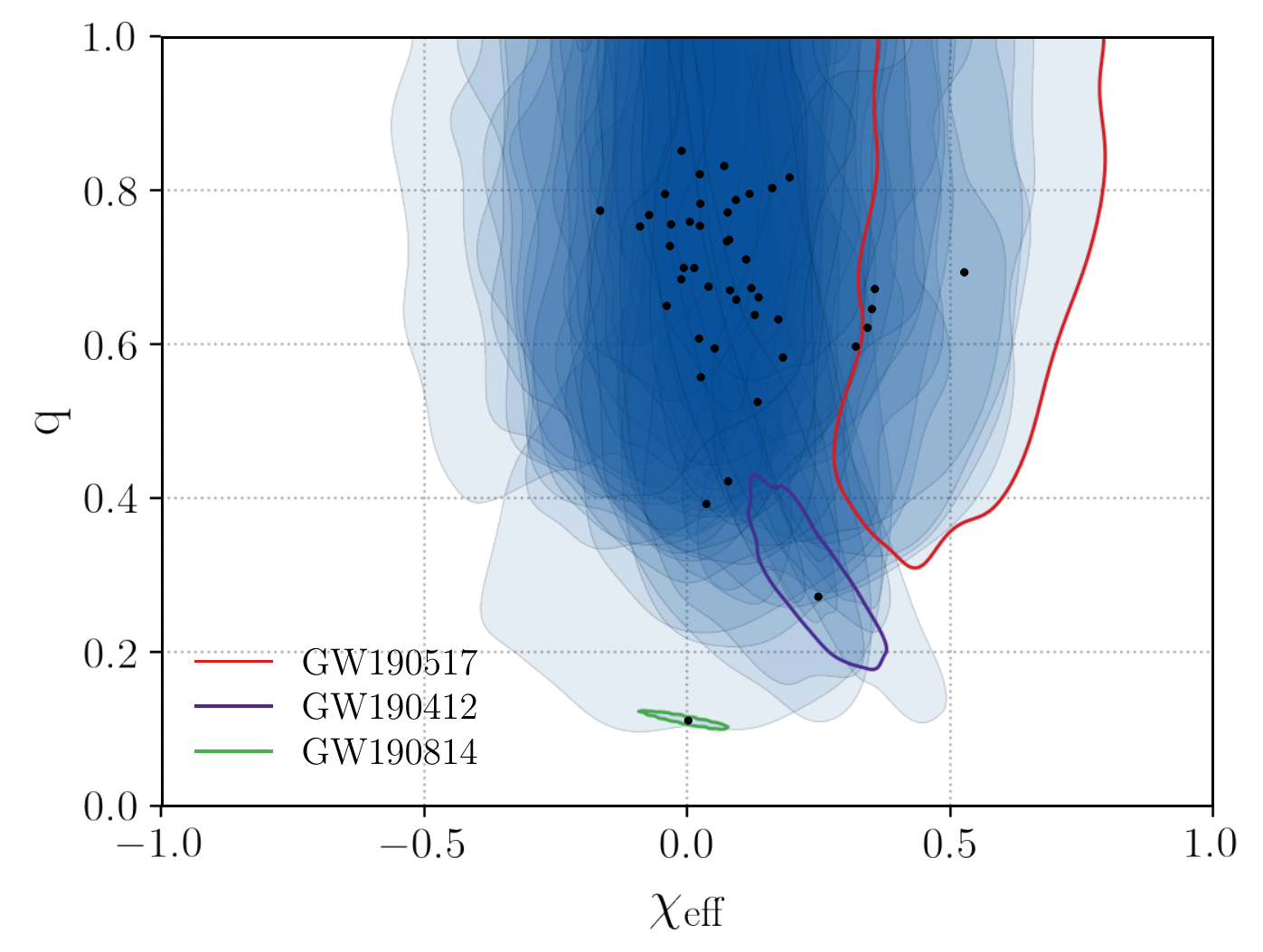}}
    \hfill
    \subfloat[\label{fig:default-posteriors-b}]{\includegraphics[width=0.48\textwidth]{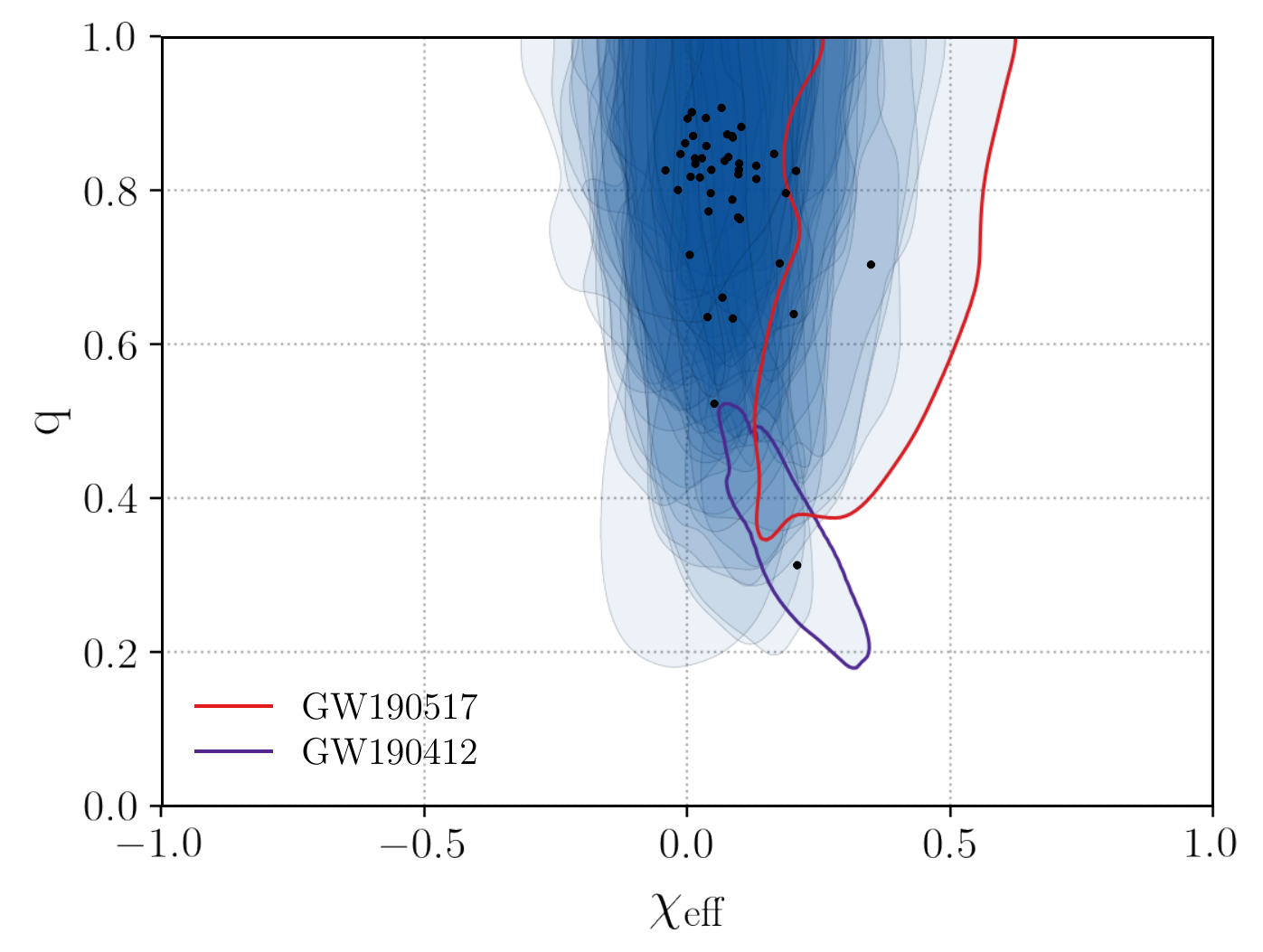}}
    \caption{
    \textit{\textbf{Left \subref{fig:default-posteriors-a}}}: Effective spin and mass ratio posteriors for the 45 BBH candidates in the LIGO/Virgo GWTC-2 catalog~\citep{gwtc2} with false alarm rates below $1\,\mathrm{yr}^{-1}$, as obtained under a default prior.
    Each shaded region gives the central 90\% credible posterior bounds for a given BBH, and black points mark the median $\chieff$ and median $q$ values for each event.
    Three individually notable events are highlighted.
    GW190412 has a precisely measured mass ratio constrained well away from unity, while GW190517 likely possesses a large, positive $\chieff$.
    Finally, GW190814 exhibits an extreme mass ratio, with a secondary that may either be a massive neutron star or a very light black hole.
    Due to the unknown nature of GW190814, this event is excluded from our analysis unless otherwise stated.
    \textit{\textbf{Right \subref{fig:default-posteriors-b}}}: Posteriors for the 44 confident BBHs (excluding GW190814) under a new population-informed prior, obtained by hierarchically fitting the BBH population assuming a Gaussian distribution of effective spins [Eq.~\eqref{eq:pchi}] and a power-law mass ratio distribution [Eq.~\eqref{eq:pq}].
    Under a population-informed prior, the ensemble of posteriors has contracted to favor smaller values of $\chieff$ and mass ratios nearer $q\sim 1$.
    }
    \label{fig:default-posteriors}
\end{figure*}

\begin{figure*}[t!]
    \centering
    \subfloat[\label{fig:default-ppc-a}]{\includegraphics[width=0.48\textwidth]{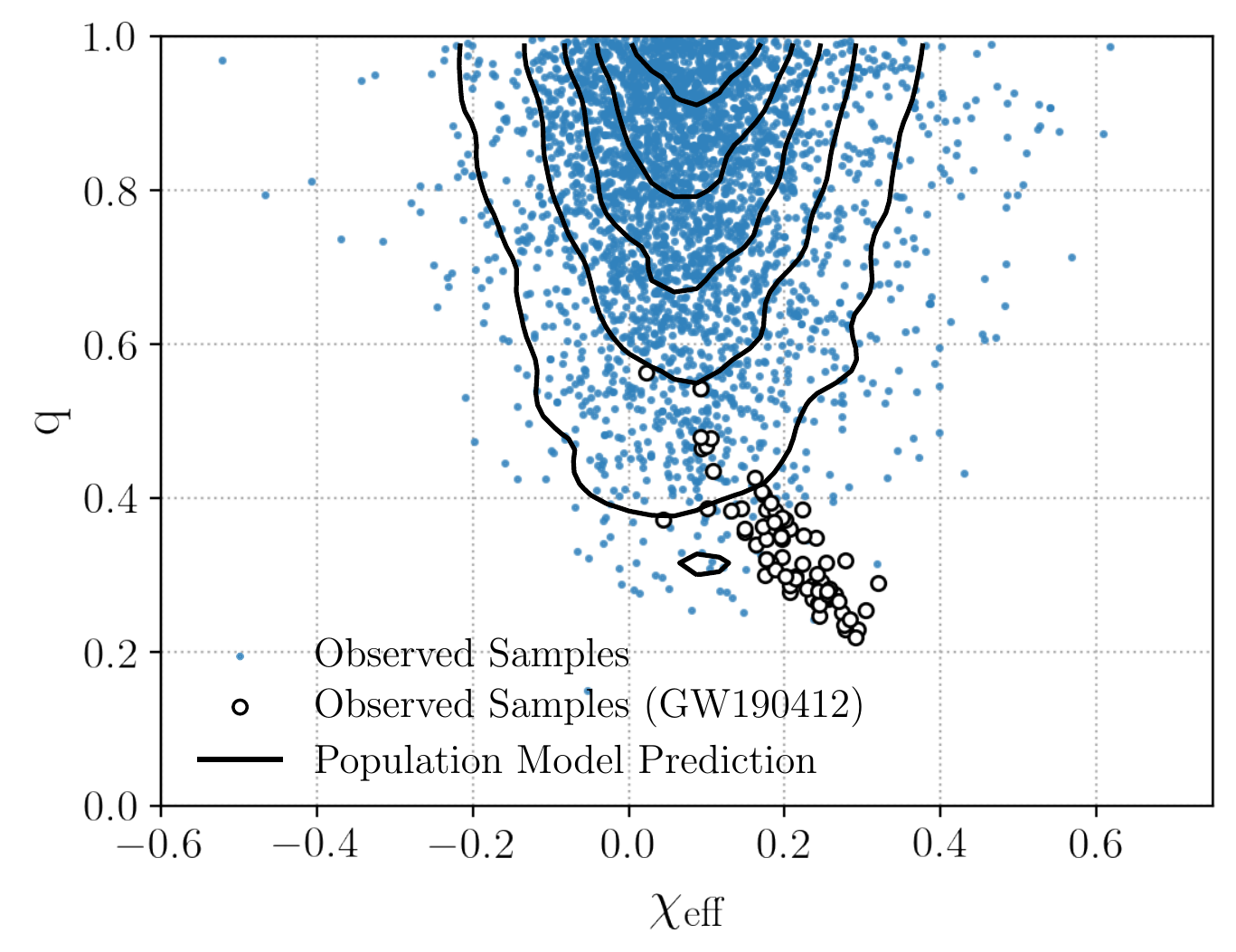}}
    \hspace{0.5cm}
    \subfloat[\label{fig:default-ppc-b}]{\includegraphics[width=0.48\textwidth]{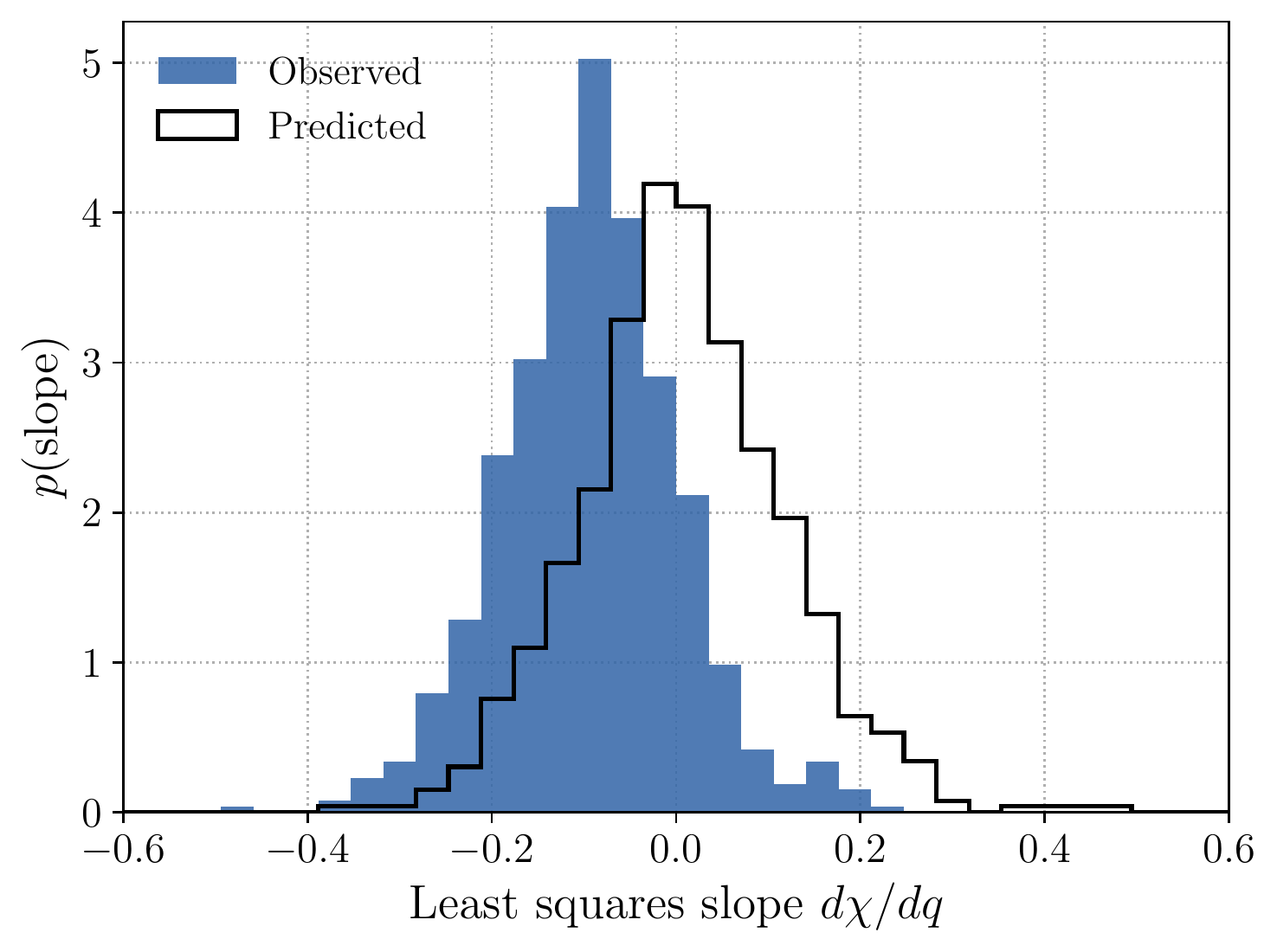}}
    \caption{
    \textit{\textbf{Left \subref{fig:default-ppc-a}}}: Locations of expected and observed detections in the $\chieff-q$ plane.
    Black contours show the expected density of future BBH detections (bounding 10\%, 30\%, 50\%, 70\%, and 90\% of expectations), given hierarchical inference on GWTC-2 under the assumption that $q$ and $\chieff$ are uncorrelated.
    Blue dots, meanwhile, show posterior draws (75 per event) from our set of BBHs in GWTC-2, reweighted to a population-informed prior as in Fig.~\ref{fig:default-posteriors-b}.
    Samples drawn from GW190412 are plotted separately as empty circles so that they can be better differentiated from the bulk population.
    \textit{\textbf{Right \subref{fig:default-ppc-b}}}: A predictive check between these expected and observed detections.
    The empty histogram shows the distribution of least-squares slopes $d\chi/dq$ under repeated draws of 44 samples from the \textit{expected} distribution of detections according to our population model.
    The blue histogram, meanwhile, shows the corresponding distribution of slopes under repeated draws from GWTC-2 posteriors (one sample per event).
    While our base population model predicts slopes centered about zero, draws from event posteriors yield a distribution of slopes offset noticeably towards negative values.
    This tension is further explored in Sect.~\ref{sec:model-fit}, in which we expand our population model to directly parametrize and measure any correlations between $q$ and $\chieff$.
    }
    \label{fig:default-ppc}
\end{figure*}

The LIGO \& Virgo second gravitational-wave transient catalog (GWTC-2) contains 44 BBH candidates with false alarm rates below one per year~\citep{gwtc2}.
For each of these 44 events, Fig.~\ref{fig:default-posteriors-a} shows the joint posterior probability on its mass ratio
    \begin{equation}
    q = \frac{m_2}{m_1}
    \end{equation}
and its effective spin~\citep{Damour2001}
    \begin{equation}
    \chieff = \frac{\chi_1 \cos t_1 + q \chi_2 \cos t_2}{1+q},
    \end{equation}
which quantifies the mass-weighted average of the two component spins when projected parallel to the binary's orbital angular momentum.
Here, $m_1$ and $m_2\leq m_1$ are the primary and secondary masses of the component black holes, $\chi_1$ and $\chi_2$ are the dimensionless component spin magnitudes, and $t_1$  and $t_2$  are the angles made by each component spin relative to the binary orbital angular momentum.
Each contour bounds the central 90\% credible region for the given event, under default parameter estimation priors (see Appendix~\ref{sec:hierachical-appendix}), and black points mark the one-dimensional median $\chieff$ and $q$ estimates for each event.

Most events in the $\chieff-q$ plane are consistent with mass ratios near unity and have effective spins clustered around $\chieff \sim 0$.
A smaller number, including the events GW190517 and GW190412, appear to have effective spins constrained away from zero.
The event GW190412 is also the first BBH to have a confidently unequal mass ratio, with $q\sim 0.3$~\citep{gw190412}.
In Fig.~\ref{fig:default-posteriors-a} we have additionally highlighted the event GW190814.
Like GW190412, GW190814 has a confidently unequal mass ratio, but its physical nature is unknown; its secondary mass $m_2 \approx 2.5\,M_\odot$ may be either a very light black hole or an extraordinarily heavy neutron star~\citep{gw190814}.
If GW190814 is presumed to be a BBH, it nevertheless remains an outlier relative to the broader BBH popluation~\citep{O3a_pop}.
We will therefore neglect GW190814 in our analysis below unless stated otherwise (we return to this event in Sect.~\ref{sec:tests}).
Also visible in Fig.~\ref{fig:default-posteriors-a} is the well-known measurement degeneracy between $\chieff$ and $q$, giving rise to the extended contours that curve down towards low $q$ and large $\chieff$~\citep{Baird2013,Ohme2013,Purrer2013,Purrer2015,Ng2018,Tiwari2018}.

In addition to these events comprising GWTC-2, independent reanalyses of Advanced LIGO and Virgo data have identified several additional BBH candidates~\citep{IAS,OGC-3}.
In our analysis below, a critical ingredient is knowledge of the selection effects governing searches for BBH events.
The LIGO and Virgo collaborations have made available a set of software injections that we will use to precisely quantify these selection effects~\citep{O3a_pop,injections}; see Appendix~\ref{sec:hierachical-appendix}.
In order to ensure self-consistency with our sample of BBHs, we will continue to use only those binaries among GWTC-2.
While this paper was in preparation, an updated GWTC-2 catalog (GWTC-2.1) was released with the addition of several low-significance BBH candidates, including two consistent with mass ratios $q\lesssim0.6$~\citep{gwtc-2.1}; we neglect these new candidates.

From Fig.~\ref{fig:default-posteriors-a} alone, it is difficult to draw any conclusions about the underlying \textit{population} of BBHs; the data in Fig.~\ref{fig:default-posteriors-a} are subject to selection effects and considerable measurement uncertainties, both of which confound any information that might be gleaned by eye about underlying features in this plane.
As a first step, we can refine Fig.~\ref{fig:default-posteriors-a} by invoking a simple model for the BBH population.
The posteriors in Fig.~\ref{fig:default-posteriors-a} are obtained via parameter estimation with broad, uninformative priors on $\chieff$ and $q$.
By hierarchically measuring the BBH population, we can reweight each posterior to a new population-informed prior, leveraging the ensemble of events to help us more accurately identify the properties of any one individual system.
For the time being, we will assume that mass ratios and effective spins are \textit{uncorrelated}, describing the population distribution of mass ratios via a power law,
    \begin{equation}
    p(q|m_1,\gamma) \propto q^{\gamma},
    \label{eq:pq}
    \end{equation}
with $m_\mathrm{min}/m_1 \leq q \leq 1$, and the distribution effective spins as a Gaussian~\citep{Roulet2019,Miller2020},
    \begin{equation}
    p(\chieff|\mu_\chi,\sigma_\chi) \propto \exp\left[ -\frac{(\chieff-\mu_\chi)^2}{2\sigma_\chi ^2}\right],
    \label{eq:pchi}
    \end{equation}
truncated on the interval $-1 \leq \chieff \leq 1$.
We obtain posteriors on the parameters governing these distributions using the \textsc{emcee} Markov Chain Monte Carlo sampler~\citep{emcee}.
Alongside $q$ and $\chi_\mathrm{eff}$, we also hierarchically measure the ensemble distribution of primary masses and BBH redshifts.
We assume that primary masses are distributed as a power law with a possible Gaussian peak, and a merger rate per comoving volume that evolves as $(1+z)^\kappa$ with  redshift~\citep{Fishbach2018,Talbot2018,O2_pop,O3a_pop}; these models are described in Appendix~\ref{sec:hierachical-appendix}, along with details of our hierarchical inference.
Our resulting posteriors on $\mu_\chi$, $\sigma_\chi$, and $\gamma$ are shown in Appendix~\ref{sec:pe-appendix}.

With this initial fit to the BBH population, we can update our measurements of $\chieff$ and $q$ for each BBH.
Figure~\ref{fig:default-posteriors-b} illustrates these reweighted posteriors.
The incorporation of a population-informed prior yields two major effects.
First, all posteriors have contracted towards small $\chieff$, since we infer both the mean and standard deviation of the effective spin distribution to be small, with $\mu_\chi = \noEvolMean$ and $\sigma_\chi = \noEvolSigma$.
Second, since $\gamma = \noEvolBetaq$ is inferred to be positive (favoring mass ratios near unity), many sources have also shifted upwards towards $q\sim 1$.

We can now explore whether this baseline population model, with \textit{uncorrelated} spin and mass ratio distributions, is a reasonably good fit to observation.
The black contours in Fig.~\ref{fig:default-ppc-a} illustrate the \textit{expected} density of detections in the $\chi_\mathrm{eff}-q$ plane, assuming our default population model is correct.
These contours are obtained using a set of simulated BBH signals injected into LIGO and Virgo data, reweighting the successfully recovered injections to our baseline population model (see Appendix~\ref{sec:hierachical-appendix}).
The expected detections are clustered at $q\sim 1$, with $\chieff$ values spread symmetrically about the population mean at $\chieff \approx 0.05$.
The collection of points, meanwhile, shows 75 posterior draws from each of our 44 BBHs, also reweighted to a population-informed prior as in Fig.~\ref{fig:default-posteriors-b}.
To better differentiate between samples drawn from the bulk population and those from the low-$q$ event GW190412, GW190412's samples are shown as empty circles, and all other posterior samples as blue dots.

To quantify the degree of tension (if any) that may exist between observed and predicted samples in Fig.~\ref{fig:default-ppc-a}, we repeatedly generate and compare catalogs of $\chieff-q$ samples consistent with GWTC-2 against predicted catalogs of mock observations drawn from our baseline population model.
We begin by choosing a random sample $\Lambda = \{\gamma,\mu_\chi,\sigma_\chi,...\}$ drawn from our posterior on the population-level parameters.
Given this value of $\Lambda$, we reweight each BBH's posterior to the corresponding population and randomly draw a single posterior sample $\{\chi_{\mathrm{eff}},q\}$ from every reweighted posterior to yield a catalog of 44 ``Observed'' values consistent with GWTC-2.
Under this same proposed population $\Lambda$, we similarly draw a ``Predicted'' catalog of mock observations, reweighting and drawing 44 events from the set of successfully found pipeline injections.
For both the ``Predicted'' and ``Observed'' catalogs we can then compute a simple least-squares slope $d\chi/dq$ of the 44 samples in the $\chieff-q$ plane.
The white and blue histograms in Fig.~\ref{fig:default-ppc-b} illustrate the distributions of these least-squares slopes, taken over many random draws of $\Lambda$.
If our baseline population model, in which $\chieff$ and $q$ are uncorrelated, were a good descriptor of GWTC-2, then draws from our model should predict slopes consistent with observation.
The distributions in Fig.~\ref{fig:default-ppc-b}, though, exhibit a systematic offset from one another: whereas the baseline population model predicts slopes centered at zero (as it must, if $\chieff$ and $q$ are presumed independent), GWTC-2 observations preferentially yield negative slopes centered above $\langle d\chi/dq\rangle = \meanObsSlopeBase$, with observed slopes lying below predicted ones $\FracBelow$ of the time.

%%%%%%%%%%%%%%%%%%%%%%
\section{Measuring a correlation between effective spin and mass ratio}
\label{sec:model-fit}
%%%%%%%%%%%%%%%%%%%%%%

\begin{figure*}[t!]
    \centering
    \includegraphics[width=0.85\textwidth]{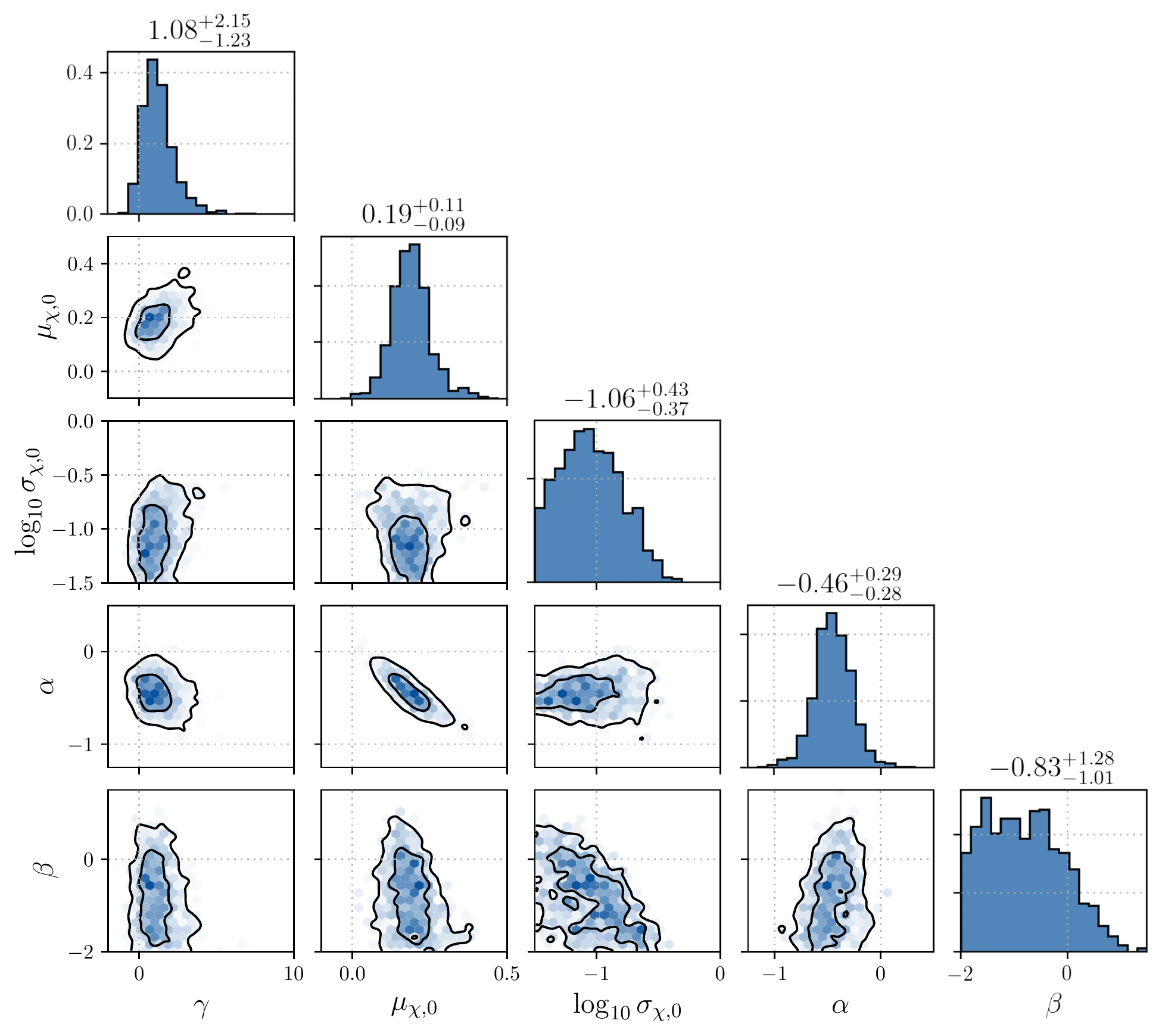}
    \caption{
    Posterior distributions on the parameters governing the mass ratio and effective spin distributions of BBHs, under a model in which the mean and standard deviation of the $\chieff$ distribution are correlated with $q$ via the parameters $\alpha$ and $\beta$ [see Eq.~\eqref{eq:pchi-evol}].
    The measurements appearing above each one-dimensional posterior correspond to median estimates and central 90\% credible uncertainties.
    At $\percentAlphaNegative$ credibility, $\alpha$ is inferred to be negative, indicating a preference for an \textit{anticorrelation} between the effective spins and mass ratios among BBHs in GWTC-2.
    Posteriors on the remaining hyperparameters used to describe the primary mass and redshift distribution are shown in Appendix~\ref{sec:pe-appendix}.
    }
    \label{fig:corner-evolution}
\end{figure*}

\begin{figure}[t!]
    \centering
    \includegraphics[width=0.43\textwidth]{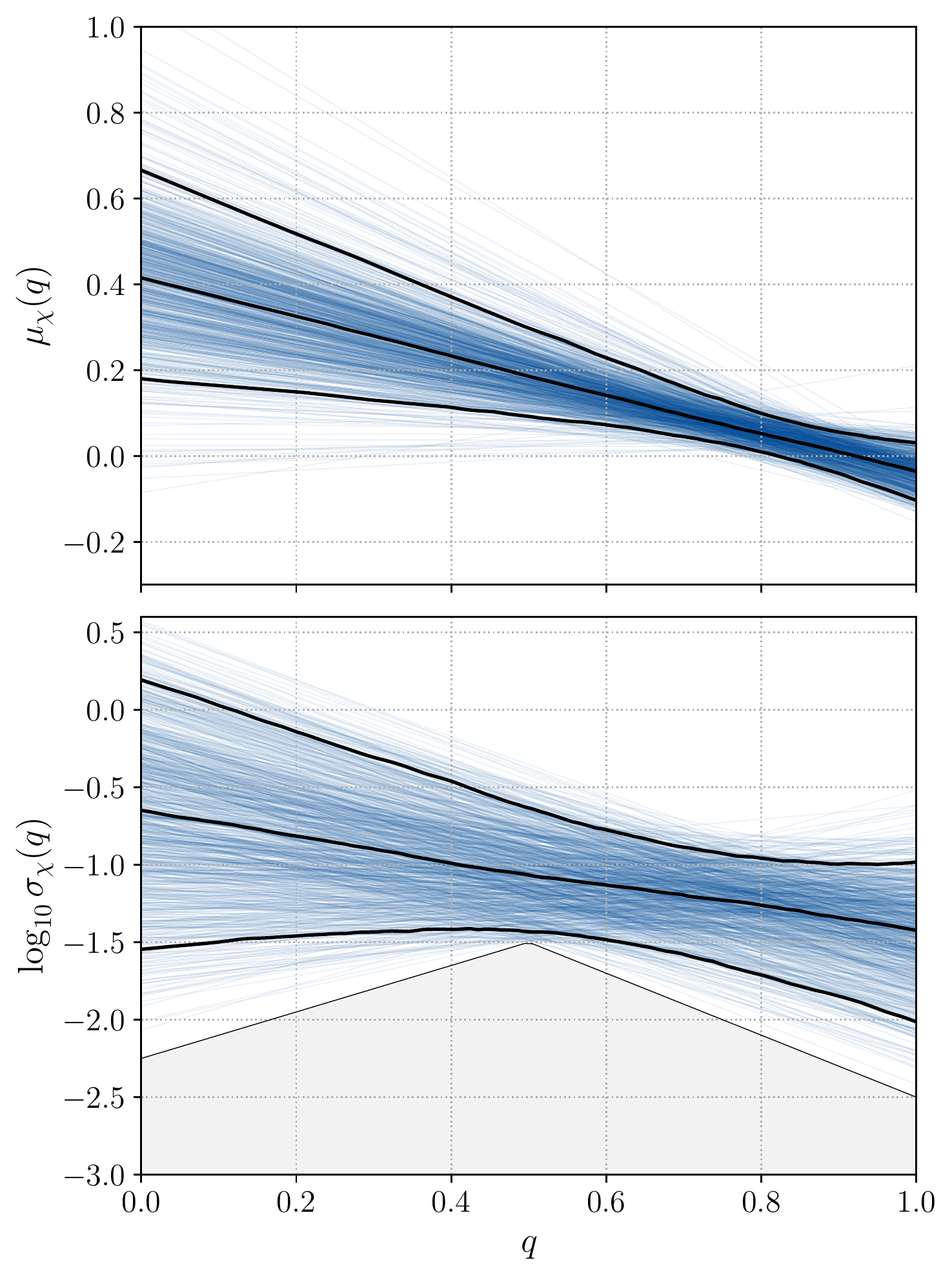}
    \caption{
    Constraints on the mean $\mu_\chi(q)$ and standard deviation $\sigma_\chi(q)$ of the $\chieff$ distribution, as a function of BBH mass ratio $q$.
    Each blue trace represents a single draw from our posterior on the population-level parameters from Fig.~\ref{fig:corner-evolution}, while black lines mark the mean and central 90\% credible bounds on $\mu_\chi(q)$ and $\sigma_\chi(q)$ at a given mass ratio.
    The mean of the $\chieff$ distribution is constrained to decrease with increasing $q$, while the data are consistent with a non-evolving width.
    In the lower panel, the shaded grey region denotes the region artificially excluded by our prior bounds on $\log\sigma_0$ and $\beta$ [see Eq.~\eqref{eq:sig-chi}].
    }
    \label{fig:traces}
\end{figure}

The baseline population model adopted in Sec.~\ref{sec:residuals}, with a power law in mass ratio (Eq.~\ref{eq:pq}) and an uncorrelated Gaussian in effective spins (Eq.~\ref{eq:pchi}), is possibly inadequate in capturing the full range of structure exhibited by BBHs in the $\chieff-q$ plane.
As explored in Fig.~\ref{fig:default-ppc}, GWTC-2 is suggestive of a tendency towards larger $\chieff$ with smaller $q$, a trend that cannot be captured with the simple population model employed so far.
Motivated by this tension, in this section we will expand our initial population model to allow for a \textit{correlation} between effective spins and mass ratios and check whether the data are informative or agnostic about the existence of such a correlation.

We continue to describe the mass ratio distribution via Eq.~\eqref{eq:pq}, but modify our $\chieff$ model such that its mean $\mu_\chi$ and log standard deviation $\log_{10}\sigma_\chi$ are now allowed to evolve linearly with $q$:
    \begin{equation}
    p(\chieff|q,\mu_{\chi,0},\sigma_{\chi,0},\alpha,\beta) \propto \exp\left[ -\frac{\big(\chieff-\mu_\chi(\mu_{\chi,0},\alpha,q)\big)^2}{2\sigma_\chi ^2(\sigma_{\chi,0},\beta,q)}\right],
    \label{eq:pchi-evol}
    \end{equation}
with
    \begin{equation}
    \mu_\chi(\mu_{\chi,0},\alpha,q) = \mu_{\chi,0} + \alpha (q-0.5)
    \label{eq:mean-chi}
    \end{equation}
and
    \begin{equation}
    \log_{10}\sigma_\chi(\sigma_{\chi,0},\beta,q) = \log_{10}\sigma_{\chi,0} + \beta(q-0.5).
    \label{eq:sig-chi}
    \end{equation}
The parameters $\alpha$ and $\beta$ govern the degree of evolution of $\chieff$ with mass ratio; confidently constraining either parameter to be non-zero would indicate that the BBH effective spin distribution shifts in location or width with increasing $q$.
Note that we choose our linear expansion in the log standard deviation rather than the standard deviation itself, in accordance with our log-uniform prior on $\sigma_\chi$ and $\sigma_{\chi,0}$ in Eqs.~\eqref{eq:pchi} and \eqref{eq:sig-chi}.

\begin{figure}[t!]
    \centering
    \includegraphics[width=0.43\textwidth]{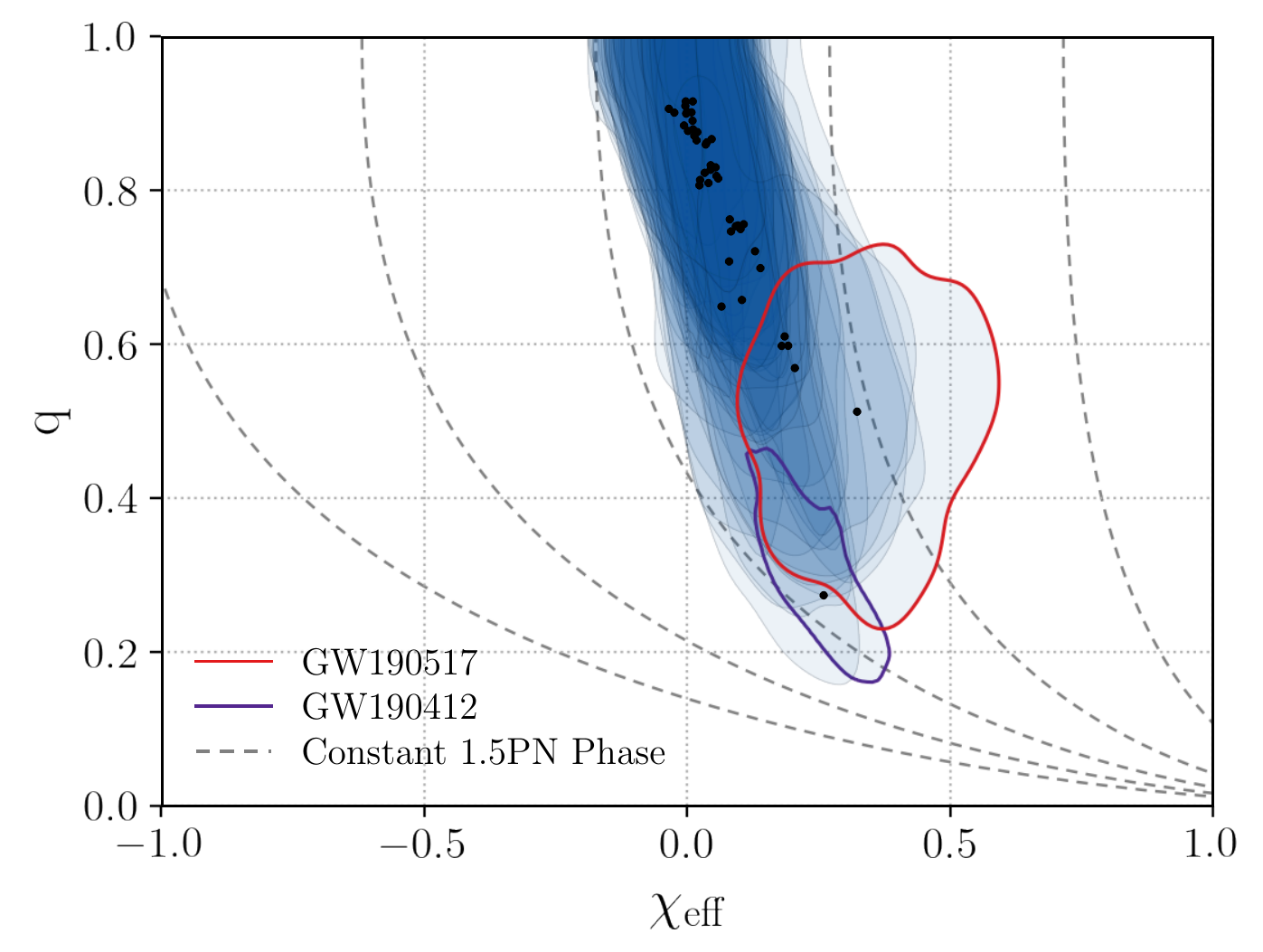}
    \caption{
    Effective spin and mass ratio posteriors for BBHs among GWTC-2, reweighted to a new population-informed prior given by our hierarchical inference of the (possibly correlated) BBH $q$ and $\chieff$ distributions.
    Under a model that allows for linear correlations between $q$ and the mean and width of the effective spin distribution, single event posteriors shift to preferentially lie along a ``best-fit line'' in the $\chieff-q$ plane.
    Notably, the high-spin event GW190517 has shifted to more unequal mass ratios, and is now constrained to the same neighborhood as the unequal-mass event GW190412.
    If BBH mass ratio and spins are correlated, it may be the case that GW190412 and GW190517 are two representatives of a shared class of BBH.
    Also shown are contours (dashed grey lines) of the 1.5PN coefficient governing the phase evolution of a gravitational-wave signal; measurement degeneracies between $\chieff$ and $q$ lie preferentially along these contours.
    }
    \label{fig:evolution-scatter}
\end{figure}

\begin{figure}[t!]
    \centering
    \includegraphics[width=0.43\textwidth]{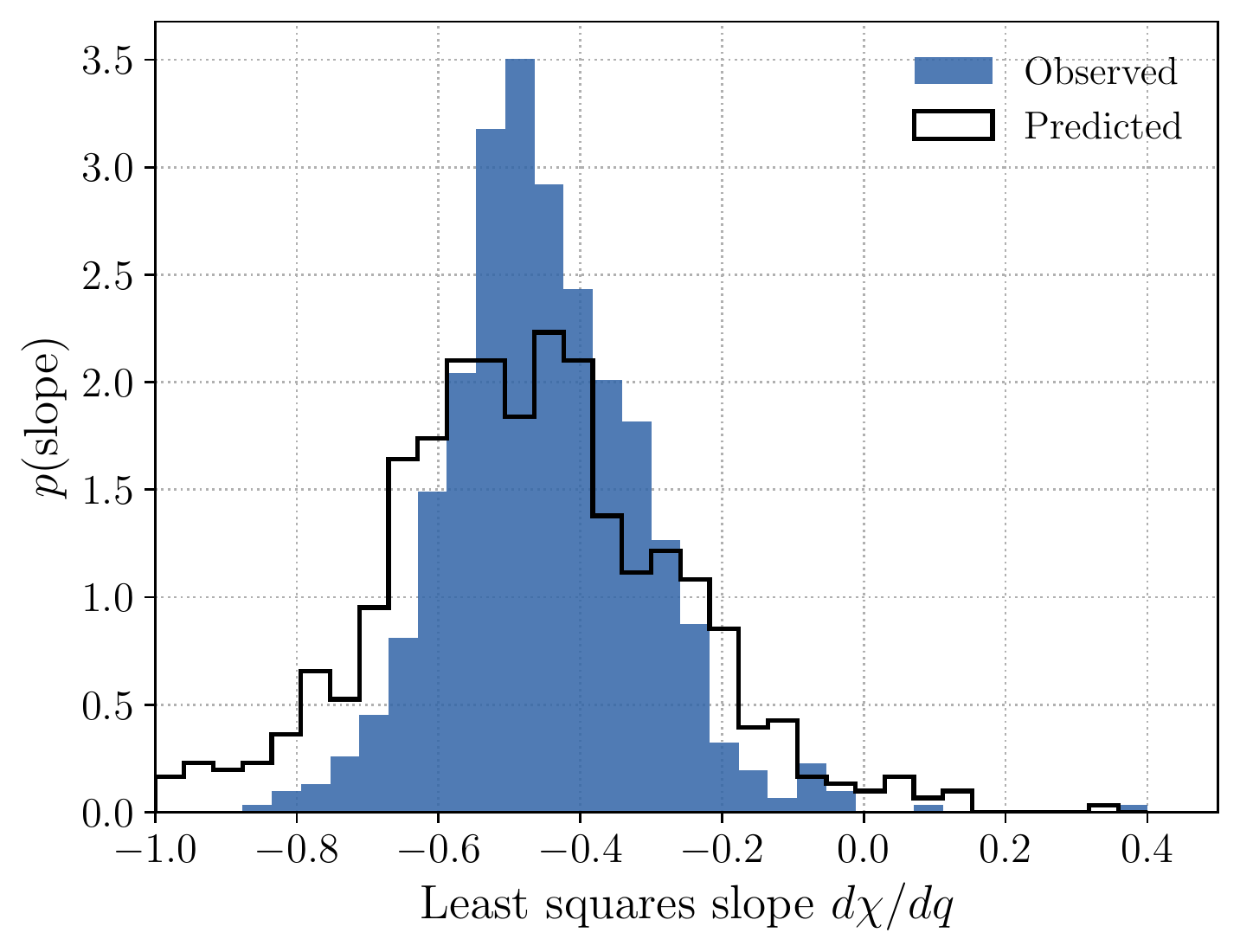}
    \caption{
    As in Fig.~\ref{fig:default-ppc-b}, a comparison of least-squares slopes $d\chi/dq$ of describing draws from the reweighted posteriors of observed BBHs to those slopes predicted by our population model that allows for $\chieff-q$ correlations.
    As shown in Fig.~\ref{fig:default-ppc}, a standard population model in which $q$ and $\chieff$ are uncorrelated yields a systematic shift between observed and predicted distributions of slopes.
    An expanded model that allows for correlations resolves this tension, yielding observed and predicted distributions centered at the same values.
    }
    \label{fig:evolution-ppc}
\end{figure}

We repeat our hierarchical analysis of the BBH events in GWTC-2, now additionally measuring the slope parameters $\alpha$ and $\beta$.
We again fit simultaneously for the primary mass and redshift distributions of BBHs, as described in Appendix~\ref{sec:hierachical-appendix}.
Figure~\ref{fig:corner-evolution} shows our resulting posteriors on the subset of parameters governing the spin and mass ratio distributions.
At $\percentAlphaNegative$ credibility, $\alpha$ is constrained to be less than zero, indicating that the effective spin distribution among BBHs in GWTC-2 is \textit{anti-correlated} with mass ratio, shifting towards larger $\chieff$ at lower $q$.

Whereas we confidently conclude that the mean of the $\chieff$ distribution evolves with $q$, we can say little about changes in the \textit{width} of the $\chieff$ distribution.
Although we rule out very large values of $\beta$, this parameter is still permitted to be moderately positive (a broadening $\chieff$ distribution towards large $q$), very negative (a narrowing distribution), or zero.
Note also that the lower boundary on $\beta$ as shown in Fig.~\ref{fig:corner-evolution} is set by our prior and not the data.

As another measure of significance, we compute a Bayes factor between our initial model in Sect.~\ref{sec:residuals}, with independent $\chieff$ and $q$ distributions, and our expanded model that allows for correlations between $q$ and $\chieff$, with possibly non-zero $\alpha$ and $\beta$.
Using the \textsc{Dynesty} nested sampler~\citep{Speagle2020} to compute Bayesian evidences for these two cases, we find a Bayes factor of \oddsCorrelatedVsUncorrelated in favor of the expanded model allowing for correlations (see Table~\ref{tab:evidences}, discussed further below).
We caution, though, that Bayes factors can be difficult to interpret due to their dependence on prior volume and the inability to capture any ``trials factors'' due to the many exploratory population analyses performed among the community.

Figure~\ref{fig:traces} illustrates our posterior on $\mu_\chi(q)$ and $\log_{10}\sigma_\chi(q)$ as a function of mass ratio.
In each subplot, black lines denote the median and central 90\% credible bounds on $\mu_\chi(q)$ and $\log_{10}\sigma_\chi(q)$, while light blue traces show the result of individual draws from our posterior on $\{\mu_{\chi,0},\log_{10}\sigma_{\chi,0},\alpha,\beta\}$.
Again, $\mu_\chi$ confidently exhibits evolution with  mass ratio, with the 90\% credible lower bound on $\mu_\chi$ at $q=0$ constrained to sit above the 90\% credible \textit{upper} bound at $q=1$.
Our best measurement of $\mu_\chi(q)$ occurs at $q\approx0.8$; the fact that we center Eq.~\eqref{eq:mean-chi} at $q=0.5$ and not $0.8$ is the source of the degeneracy between $\mu_{\chi,0}$ and $\alpha$ seen in Fig.~\ref{fig:corner-evolution} above.
In contrast, no evolution is evident for $\log_{10}\sigma_\chi(q)$: a horizontal line can be fit inside the 90\% credible bounds shown, consistent with the fact that we do not exclude $\beta = 0$.
Note that, since our posterior on $\beta$ is bounded by our prior, a non-negligible fraction of the space in Fig.~\ref{fig:traces} is excluded \textit{a priori}.
The shaded grey area shows this region that is artificially excluded by our priors on $\log_{10}\sigma_0$ and $\beta$.

Reweighting the posterior of each BBH via an updated population prior from our correlated $\chieff-q$ model yields the results shown in Fig.~\ref{fig:evolution-scatter}.
Relative to the posteriors obtained via standard parameter estimation (Fig.~\ref{fig:default-posteriors-a}) and those obtained by reweighting to a standard uncorrelated population model (Fig.~\ref{fig:default-posteriors-b}), the reweighted posteriors in Fig.~\ref{fig:evolution-scatter} exhibit a significant degree of ``shrinkage,'' shifting to lie predominantly along the ``best-fit'' line characterized by $\alpha \approx -0.5$.
The event GW190517, in particular, changes character considerably under our updated population prior.
Under standard parameter estimation priors, GW190517 is a possible outlier in spin, with the largest $\chieff$ measured among GWTC-2, but otherwise is consistent with a fairly typical mass ratio~\citep{gwtc2}.
Under a population model that allows for correlated mass ratios and spins, however, the posterior of GW190517 shifts to favor significantly lower $\chieff$ and $q$, identifying GW190517 as another confidently unequal mass event like GW190412.
Moreover, neither GW190517 nor GW190412 are clear outliers relative to the rest of the BBH population; both are quite consistent with the same linear trend favored by the other BBH observations.

Although the BBHs in GWTC-2 favor a $\chieff$ distribution that evolves mass ratio, does this expanded model actually  resolve the predictive tension discussed in Fig.~\ref{fig:default-ppc}?
We repeat the predictive exercise discussed in Sect.~\ref{sec:residuals}, repeatedly drawing a hyperparamter sample $\Lambda = \{\gamma,\mu_{\chi,0},\sigma_{\chi,0},\alpha,\beta,...\}$, generating ``Observed'' catalogs from our reweighted posteriors and ``Predicted'' catalogs from the population model, and recording their least-squares slopes.
The resulting distributions are shown in Fig.~\ref{fig:evolution-ppc}.
The expectation values over these distributions are now in far better agreement, with the ``Observed'' and ``Predicted'' slopes having sample means of $\langle d\chi/dq\rangle = \meanObsSlopeEvol$ and $\meanMockSlopeEvol$, respectively.

%TC:ignore
%%%%%%%%%%%%%%%%%%%%%%
\section{Spurious sources of apparent correlation?}
\label{sec:tests}
%%%%%%%%%%%%%%%%%%%%%%

An anti-correlation of BBH effective spins with mass ratio is surprising, given current theories of compact binary formation and evolution.
In this section, we therefore detail a number of checks to bolster our confidence that this anti-correlation is a real observed feature of our data, rather than a spurious effect  due to uncontrolled systematics.

\begin{figure}
    \centering
    \includegraphics[width=0.475\textwidth]{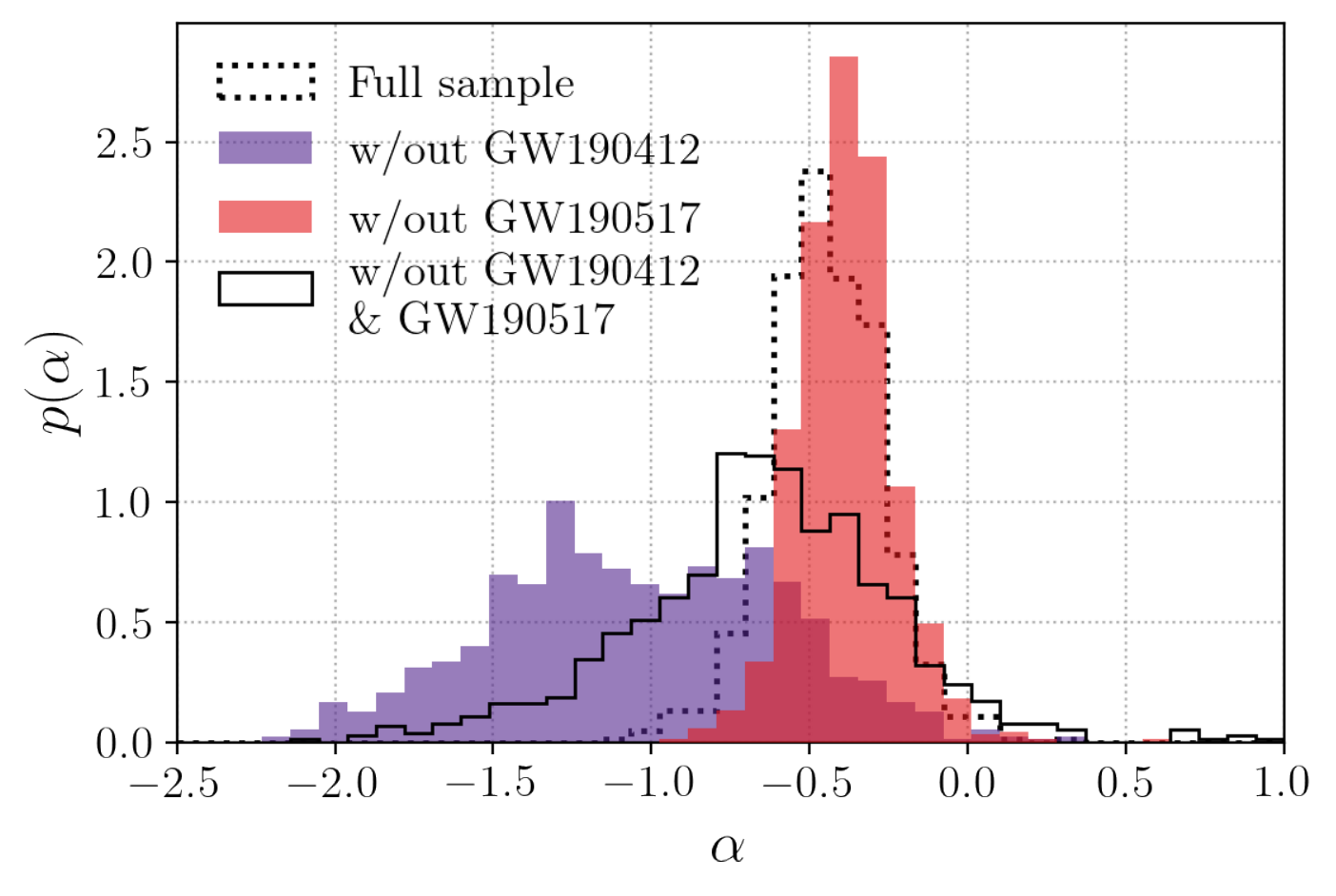}
    \caption{
    Posteriors on the parameter $\alpha$ governing the evolution of $\mu_\chi$ with mass ratio when our hierarchical inference is repeated with GW190412 and/or GW190517 left out of our sample.
    When one or the other event is excluded, $\alpha$ remains confidently negative.
    This holds true even when \textit{both} events are excluded, although with  reduced statistical significance; in this case, $\alpha$ is constrained below zero at $\percentAlphaNegativeNoOhFourTwelveOrOhFiveSeventeen$ credibility.
    }
    \label{fig:out-in}
\end{figure}

First, we explore whether our confident measurement of $\alpha<0$ can be traced to any one event.
If so, this would not necessarily indicate that our results are spurious, but would suggest that the event in question is an outlier among the BBH population, or possibly that an error or bias is present in the data associated with this event.
We repeat our hierarchical inference of the correlated $\chieff-q$ model in Sect.~\ref{sec:model-fit} three additional  times, excluding GW190412, excluding GW190517, and excluding both events together.
The resulting marginalized posteriors on $\alpha$ are shown in Fig.~\ref{fig:out-in}.
The observed anti-correlation between $\chieff$ and $q$ remains robust against the exclusion of either event.
Excluding GW190517 leaves our result effectively unchanged, with $\alpha<0$ at $\percentAlphaNegativeNoOhFiveSeventeen$ credibility.
The same is true of GW190412, whose exclusion gives $\alpha<0$ at $\percentAlphaNegativeNoOhFourTwelve$ credibility.
Finally, while excluding \textit{both} GW190412 and GW190517 somewhat weakens our conclusions, in this case $\alpha$ is still constrained to be less than zero at $\percentAlphaNegativeNoOhFourTwelveOrOhFiveSeventeen$ credibility.
Therefore, although these two events notably aid in the precision with which $\alpha$ is measured, the preference for negative $\alpha$ is a feature of the broader BBH population.
We note that, when GW190412 is excluded, \textit{more negative} values of $\alpha$ are permitted by the data, suggesting that, rather than driving the increase in $\chieff$ with lower $q$, the tight parameter estimates for GW190412 appear instead to anchor any evolution to somewhat shallower values.
This behavior is further explored in Appendix~\ref{sec:most-important-events}, where we quantify which events most strongly prefer or most strongly resist an anti-correlation between $q$ and $\chieff$.

\begin{figure*}
    \centering
    \subfloat[\label{fig:gw190814-a}]{\includegraphics[width=0.512\textwidth]{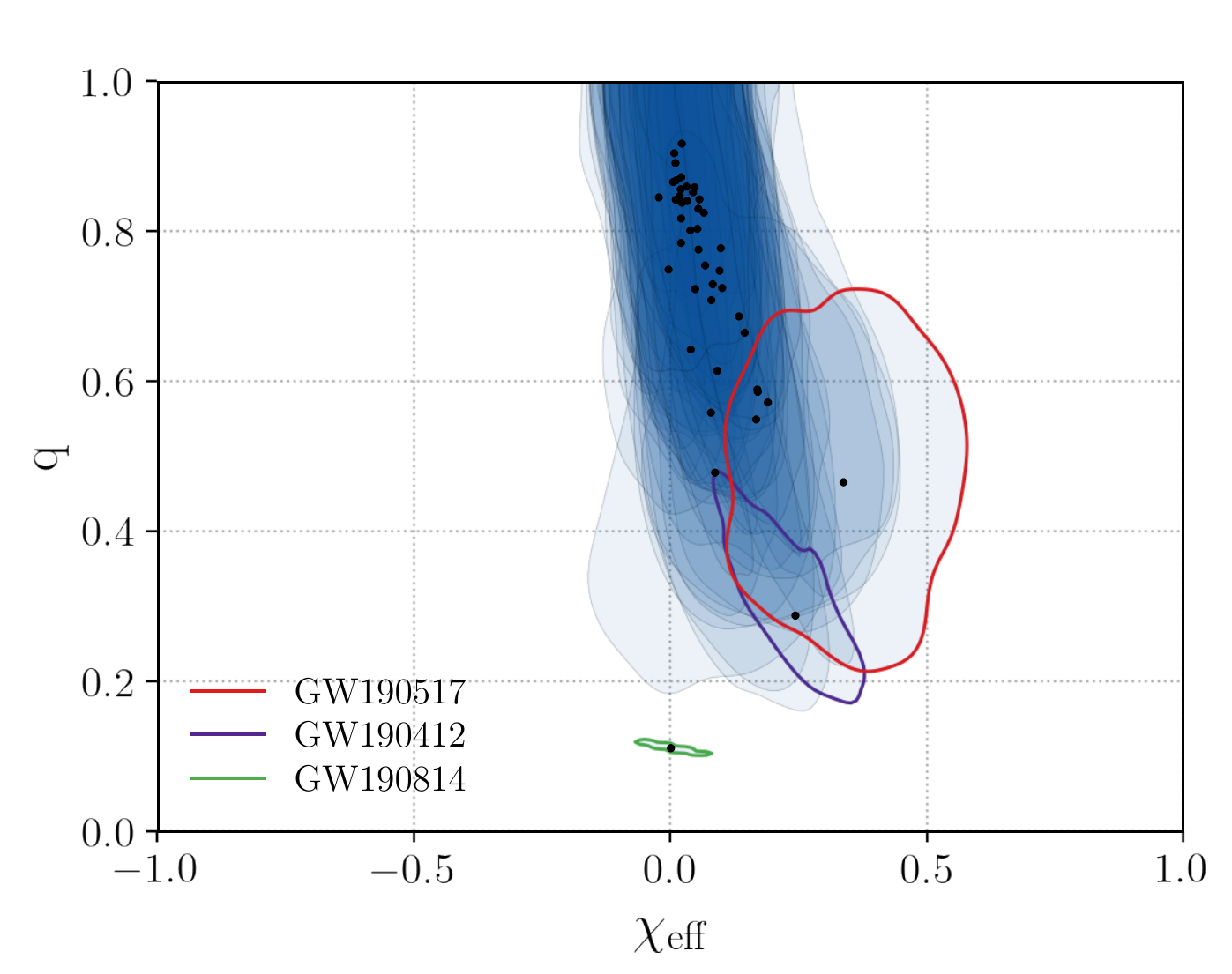}}
    \hfill
    \subfloat[\label{fig:gw190814-b}]{\includegraphics[width=0.475\textwidth]{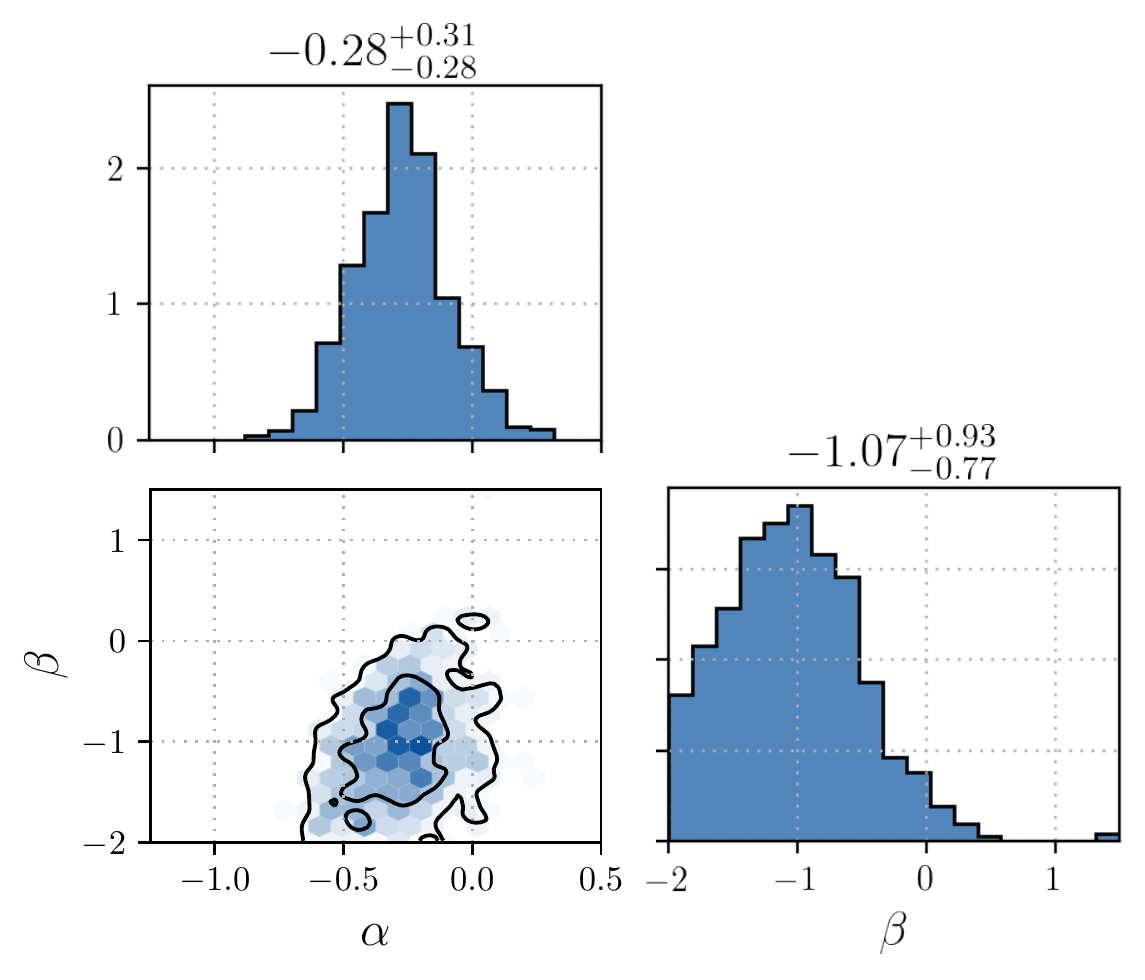}}
    \caption{
    Hierarchical inference results and reweighted single-event posteriors when the outlier event GW190814 is \textit{included} in our BBH sample.
    The inclusion of GW190814 diminishes our confidence that the mean of the $\chieff$ distribution evolves with mass ratio.
    Instead, this event leads us to conclude that the \textit{width} of the $\chieff$ distribution likely increases with decreasing $q$.
    Comparing the left-hand panel to the reweighted posteriors in Fig.~\ref{fig:evolution-scatter} (in which GW190814 is excluded from the fit), the bulk of the BBH events still cluster along a best fit line, but, in order to additionally accommodate the precisely-measured properties of GW190814, exhibit with an increasing degree of scatter about this line at smaller mass ratios.
    }
    \label{fig:gw190814}
\end{figure*}

If, on the other hand, we \textit{add in} GW190814, our conclusions do not change qualitatively.
Due to its unknown source classification, we have so far excluded GW190814 from the set of BBHs informing our hierarchical inference~\citep{gw190814,O3a_pop}.
If we repeat our analysis but now include GW190814 among our sample, we obtain the reweighted population shown in Fig.~\ref{fig:gw190814-a}, while Fig.~\ref{fig:gw190814-b} shows the updated posteriors on $\alpha$ and $\beta$ when GW190814 is present.
Given the very precise measurements of GW190814's mass ratio and effective spin, this event does not shift noticeably under a population-informed prior.
Its presence, though, doesn't override the tendency of the other BBHs to prefer anticorrelated $\chieff$ and $q$.
We still confidently infer $\alpha<0$, although the strength of this measurement, now at $\percentAlphaNegativeWithOhEightFourteen$ credibility, is reduced.
GW190814's primary effect, however, is to \textit{broaden} the $\chieff$ distribution towards lower $q$; we now infer that $\beta<0$ at $\percentBetaNegativeWithOhEightFourteen$ credibility (although the exact significance will depend on our lower prior bound on $\beta$).
Qualitatively, since all other BBHs force $\alpha<0$, the only way for our population model to simultaneously accommodate GW190814 is also to broaden at very low $q$.
This effect can be seen when comparing the reweighted posteriors in Figs.~\ref{fig:evolution-scatter} and \ref{fig:gw190814-a}: whereas the median measurements (black points) in Fig.~\ref{fig:evolution-scatter} obey a tight linear correlation, those in Fig.~\ref{fig:gw190814-a} exhibit visibly increased scatter as we move towards small $q$.

It could alternatively be the case that biases due to our choice of mass model or systematic uncertainties in the Advanced LIGO \& Virgo selection function yield an artificial preference for  $\alpha <0$.
The fact that our mass ratio distribution $p(q|m_1)$ is specified as \textit{conditional} on $m_1$ implies that our conclusions regarding $q$, and hence also our conclusions regarding correlations between $q$ and $\chieff$, are possibly impacted by our choice for $p(m_1)$.
Our results, though, behave robustly against different choices for the form of $p(m_1)$.
So far, we have shown results obtained while assuming $p(m_1)$ is a power law with a Gaussian peak \citep[the \textsc{Power Law+Peak} model of][]{O3a_pop}.
If we instead adopt a broken power-law form for $p(m_1)$, our conclusions are virtually unchanged.
Similarly, if we instead assume that $p(q)$ is itself a Gaussian rather than a power law, we obtain a consistent estimate of $\alpha$.
As discussed in Appendix~\ref{sec:hierachical-appendix}, meanwhile, we account for selection effects in O3 by using the results of an actual injection campaign into Advanced LIGO \& Virgo data~\citep{gwtc2,O3a_pop,injections}.
Analogous injections are not available for the O1 and O2 observing runs, however; for these observing runs we instead estimate the LIGO \& Virgo selection function using a publicly-available set of mock events that pass a semi-analytic signal-to-noise ratio cut~\citep{pop_data_release}.
It is conceivable that this semi-analytic calculation biases us in unexpected ways.
To check this possibility, we have verified that our results hold when performing hierarchical analysis using only those events detected in O3a, thereby avoiding any possibility of bias due to the O1 and O2 selection function.

Another possible concern is the well-known fact that measurements of individual events' $\chieff$ and $q$ exhibit a large degree of correlation~\citep{Cutler1994,Poisson1995,Baird2013,Ohme2013,Purrer2013,Purrer2015,Ng2018,Tiwari2018}.
Both the effective spin and mass ratio appear in the 1.5PN coefficient in the post-Newtonian expansion of a BBH's phase evolution.
It is this coefficient that is most readily extracted from a gravitational-wave observation, and so the resulting individual measurements of $\chieff$ and $q$ are quite degenerate with one another, particularly in the case of low-mass systems for which the inspiral (rather than merger and ringdown) contributes the bulk of the observed signal-to-noise.
This degeneracy can be seen in Fig.~\ref{fig:default-posteriors}; if examined closely, the posteriors for several individual events take the form of extended arcs that curve down and to the right, towards large $\chieff$ and small $q$.
Since this degeneracy acts in the \textit{same} direction as the $\chieff-q$ anti-correlation we identify, one might wonder if our observation of an anti-correlation is just a relic of these measurement degeneracies, rather than an intrinsic feature of the underlying BBH population.

If the Bayesian parameter estimation of individual gravitational-wave events is unbiased, hierarchical Bayesian inference is robust against measurement degeneracies.
Large degeneracies may well inflate our \textit{uncertainties} on the properties of the underlying BBH population, but will not lead to an erroneous identification of some spurious feature or correlation.
In practice, though, we cannot be absolutely confident in the complete and unbiased coverage of parameter estimation.
The properties of the massive BBHs GW151226 and GW190521, for example, are still under debate due to the possible existence of additional posterior modes identified upon reanalysis of LIGO/Virgo data~\citep{gw151226,gw190521,Nitz2021,Estelles2021,Mateu2021,Chia2021}.
Moreover, it is in principle possible for small but systematic errors in our estimates of parameter estimation \textit{priors}, a critical ingredient in hierarchical inference, to have an outsized cumulative impact on our population-level results.

\begin{figure*}
    \centering
    \subfloat[\label{fig:injection-a}]{\includegraphics[width=0.485\textwidth]{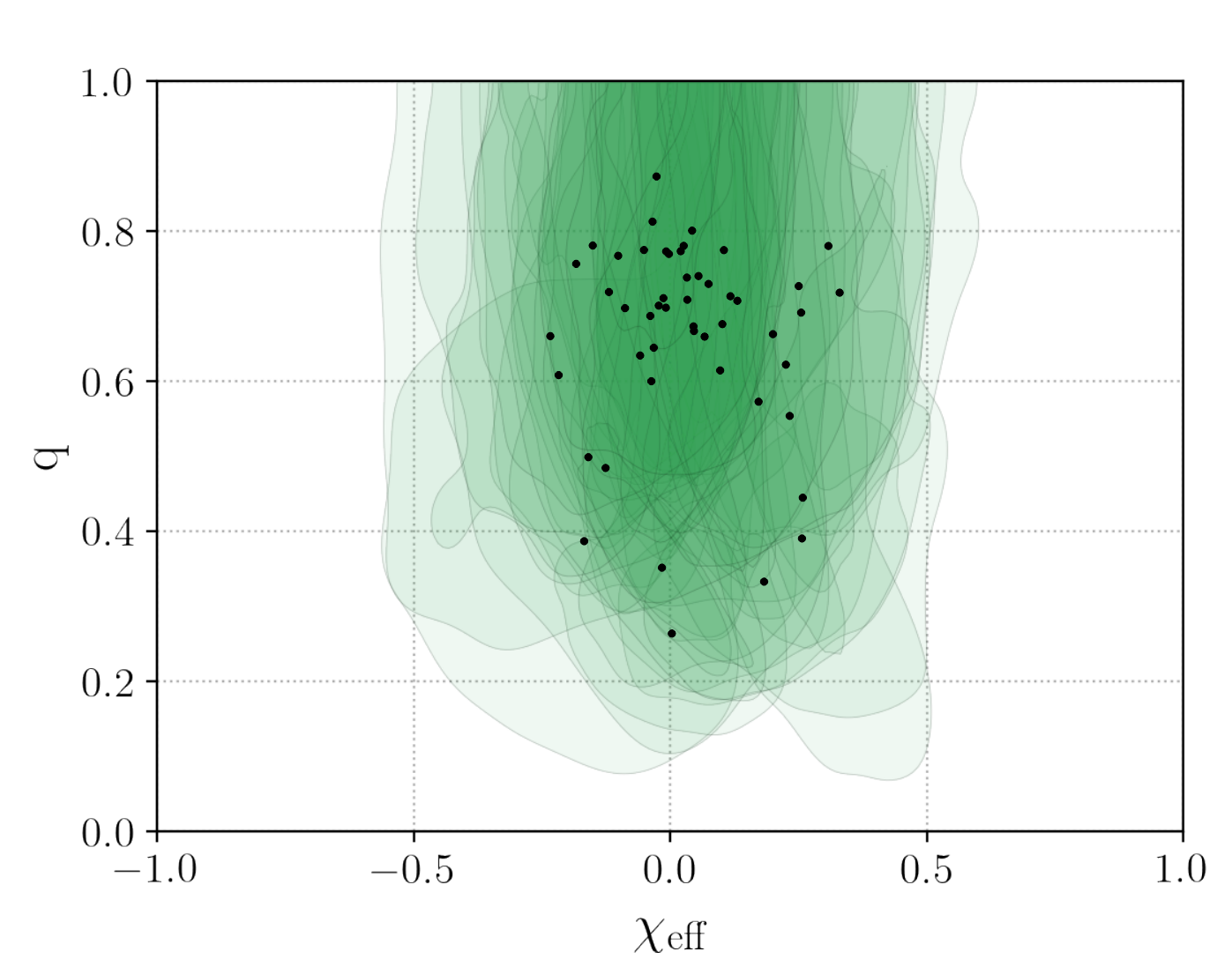}}
    \hfill
    \subfloat[\label{fig:injection-b}]{\includegraphics[width=0.48\textwidth]{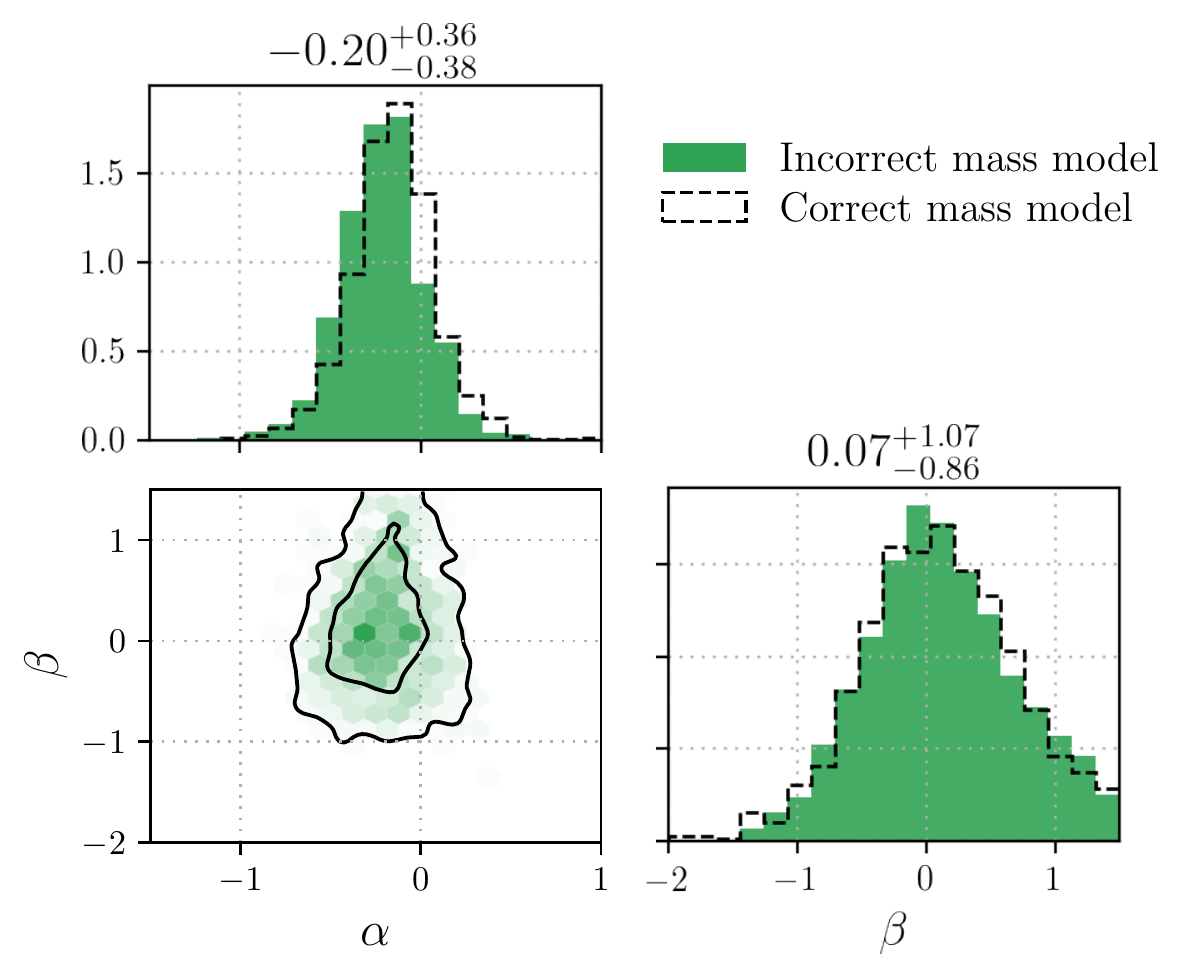}}
    \caption{
    \textit{\textbf{Left \subref{fig:injection-a}}}:
    Parameter estimation posteriors for a catalog of mock BBH detections.
    Each contour encloses the central 90\% credible region for a given event, while black points mark one-dimensional medians on $q$ and $\chieff$.
    These mock detections are drawn from a population with \textit{no} intrinsic correlations between $q$ and $\chieff$, but with parameters otherwise consistent with those of GWTC-2.
    Our resulting posteriors display the same key features as those of real BBH detections in Fig.~\ref{fig:default-posteriors-a}, including measurement degeneracy between $q$ and $\chieff$.
    \textit{\textbf{Right \subref{fig:injection-b}}}:
    Inferred posteriors on the parameters governing correlations between $q$ and $\chieff$ [see Eqs.~\eqref{eq:mean-chi} and \eqref{eq:sig-chi}] when performing hierarchical analysis on the mock catalog.
    Our results are consistent with $\alpha = \beta = 0$, correctly reflecting the fact that our underlying injections possess no intrinsic correlations between $q$ and $\chieff$.
    As discussed further in the text, we further test the stability of this result by deliberately analyzing this population with an incorrect mass model (power law with a Gaussian peak; green).
    Results obtained using the \textit{correct} mass model from which injections are drawn (broken power law) are shown via the dashed histograms.
    }
    \label{fig:injection}
\end{figure*}

As a purely qualitative check, we can see to what degree our population-level $\chieff-q$ anti-correlation and the $\chieff-q$ measurement degeneracies lie in the same direction.
The 1.5PN phase correction may be written $\Psi_{1.5} = \left(\pi \mathcal{M}_c f\right)^{-2/3} \psi_{1.5}$, with the coefficient~\citep{Cutler1994,Poisson1995,Baird2013,Ng2018}
    \begin{equation}
    \psi_{1.5} = \eta^{-3/5} \left[
        \frac{(113-76\eta)}{128}\chieff - \frac{76 \eta \delta}{128} \chi_a - \frac{3\pi}{8}
        \right].
    \label{eq:pn-phase}
    \end{equation}
Here, $\mathcal{M}_c$ is the binary chirp  mass, $\eta$ is the symmetric mass ratio, $\delta = (m_1-m_2)/(m_1+m_2)$, and $\chi_a = (\chi_1 \cos t_1 - \chi_2 \cos t_2)/2$.
The dashed lines in Fig.~\ref{fig:evolution-scatter} trace contours of equal $\psi_{1.5}$, assuming that $\chi_a = 0$.
It is along these contours that measurement degeneracies between $q$ and $\chieff$ preferentially lie.
Note that different choices for $\chi_a$ will change the exact contours drawn but that these differences are small, since the influence of $\chi_a$ in Eq.~\eqref{eq:pn-phase} is suppressed by a factor of $\delta$ relative to the term involving $\chieff$.
If our hierarchical inference were contaminated by measurement degeneracies, we expect that our analysis would favor a correlation lying parallel to these constant-in-phase contours.
The slope $\alpha \approx -0.45$ favored by the BBH population, however, differs from these contours at the mass ratios $q\gtrsim 0.6$ where most detections lie.

This check is purely qualitative, however, and may be affected by the fact that different events lie along different 1.5PN contours, as well as the fact that high mass events have better measured total masses than chirp masses.
In Appendix~\ref{sec:most-important-events}, we quantify Bayesian evidences for each event in our sample between two fixed populations: one that includes a $q-\chi_\mathrm{eff}$ anti-correlation and one that excludes it.
Of the four events that most strongly prefer a correlated population model, none exhibit the sweeping 1.5PN degeneracies discussed here, offering further evidence that measurement degeneracies are not spuriously driving our measurement of negative $\alpha$.

As another more concrete test, we verify that measurement degeneracies do not reproducibly confound our analysis by performing an end-to-end injection, recovery, and hierarchical inference of a mock population of BBH events.
We randomly draw a large set of BBHs from a broad reference population (see Appendix~\ref{sec:inj-appendix} for additional details), down-selecting to $5\times10^4$ ``found'' events that have a matched-filter network signal-to-noise ratio $\rho\geq 10$ across LIGO-Hanford, LIGO-Livingston, and Virgo.
As we are concerned only with $\chieff$, for simplicity we work only with aligned spins, using the \textsc{IMRPhenomD} waveform model~\citep{Husa2016,Khan2016}.
From our large sample of found injections, we randomly draw $N=50$ to comprise our ``observed'' catalog on which we will perform parameter estimation and hierarchical inference.
In drawing this catalog, we assign draw probabilities such that our injected population has $\mu_0 = 0.05$, $\sigma_0 = 0.15$, $\alpha = 0$, and $\beta = 0$, with \textit{no} intrinsic correlation between $\chieff$ and $q$.
Additionally, we deliberately introduce a mismatch between the injected distribution of primary masses and the distribution assumed on recovery; while primary masses are drawn from a broken power law, we will fit this population  using the same power law and Gaussian peak mixture adopted above.

We perform parameter estimation on each of these 50 events using \textsc{Bilby}~\citep{Ashton2019,Romero-Shaw2020} in conjunction with the \textsc{Dynesty}~\citep{Speagle2020} nested sampler.
Figure~\ref{fig:injection-a} shows the posteriors we recover from our injections under a default parameter estimation prior.
This ensemble of posteriors displays (partially by design) many of the same qualitative features of GWTC-2 seen above in Fig.~\ref{fig:default-posteriors-a}, with most events clustered near small $\chieff$ and moderately large $q$, a handful of posteriors whose medians are displaced towards positive $\chieff$, and several events exhibiting the curving degeneracy characteristic of joint $\chieff$ and $q$ measurements.
Finally, we use the set of resulting posterior samples to hierarchically analyze our injected population, employing the population model described in Sect.~\ref{sec:model-fit} to fit for any correlations between $\chieff$ and $q$, obtaining the posteriors on $\alpha$ and $\beta$ shown in Fig.~\ref{fig:injection-b}
We recover $\mu_0$ and $\sigma_0$ estimates consistent with our injected values, finding $\mu_{\chi,0} = \injMean$ and $\log_{10}\sigma_{\chi,0} = \injSigma$ (medians and central 90\% credible uncertainties).
More importantly, our results are consistent with \textit{no correlations} between effective spin and mass ratio.
Our posteriors allow quite comfortably for $\alpha=0$ and $\beta=0$, which lies on a contour enclosing $\injQuantileAtOrigin\%$ of the probability in the $\alpha-\beta$ plane.
Hence we correctly conclude that our injected population exhibits no correlations between $\chieff$ and $q$.
As noted above, we deliberately perform our hierarchical inference with an incorrect mass model.
For completeness, the dashed histograms in Fig.~\ref{fig:injection-b} show the marginal posteriors obtained on $\alpha$ and $\beta$ if we instead fit our injection set with the \textit{correct} mass model (a broken power law in primary mass).
Our conclusions regarding $\alpha$ are effectively unchanged between these two cases.

This injection case study does not, of course, serve as proof that there are no unknown sources of bias in GWTC-2.
It does, however, demonstrate that a false-positive identification of $\chieff$ and $q$ does not readily appear when hierarchically analyzing fully realistic BBH posteriors, particularly those posteriors that exhibit measurement degeneracy between mass ratio and effective spin.

It is possible that our analysis is affected by other systematic biases due to imperfect detector calibration and/or biases in the waveform models used for parameter estimation.
Both possibilities, though, are unlikely to give rise to the observed $q-\chieff$ anti-correlation.
Parameter estimation samples used in this study have been marginalized over a frequency-dependent calibration uncertainty budget~\citep{gwtc2,O1O2-calibration,O3a-calibration}.
As discussed in Appendix~\ref{sec:hierachical-appendix}, meanwhile, we use a union of parameter estimation samples from several distinct waveform families, mitigating potential biases that might be peculiar to any one waveform model.
%TC:endignore

\section{Astrophysical Implications}
\label{sec:implications}

The physical implications of an anti-correlation between $q$ and $\chieff$ are unclear.
There are generally two approaches one might take in attempting to explain this relationship.
First, such an anti-correlation could conceivably arise from processes acting within a \textit{single} population of BBHs arising from a common formation channel.
Second, a global anti-correlation could arise if observed BBHs originate from some superposition of formation channels, with some favoring high $q$ and low $\chieff$ and others yielding high $\chieff$ with low $q$.
We comment on each of these possibilities in turn.

BBHs arising from isolated stellar binaries have spins that result from a complex interplay of angular momentum transport in stellar cores, tidal torques operating between stars, and episodes of mass transfer~\citep{Spruit2002,Gerosa2018,Qin2018,Zaldarriaga2018,Belczynski2019,Fuller2019,Bavera2020,Bavera2021}.
The details of the latter two processes depend on binary mass ratio, and so might impart a relationship between $q$ and $\chieff$.
The generally predicted relationship between $q$ and $\chieff$, however, differs from the relationship we see here observationally.
\citet{Bavera2020}, for instance, predict that BBHs originating from common envelope (CE) exhibit an increased \textit{scatter} in $\chieff$ towards positive values with increasing $q$.
\citet{Bavera2021} later identify a similar trend among binaries that undergo Eddington-limited stable mass transfer.
In our analysis, this effect would manifest as positive values for $\alpha$ and $\beta$; this possibility is ruled out at high credibility.
We do note, though, that the simulations of \citet{Bavera2021} do contain two special cases in which isolated BBHs could exhibit anti-correlated $q$ and $\chieff$.
CE binaries with very high common-envelope efficiencies and stable mass transfer binaries with super-Eddington accretion \citep[e.g. fourth row in Fig.~G.2 and third row in  Fig.~G.3 of][respectively]{Bavera2021} each predict structure in the $\chieff-q$ plane that is at least qualitatively similar to the behavior we find here.

Dynamically-assembled binaries in dense stellar clusters may also naturally exhibit a correlation between $q$ and $\chieff$ if they experience repeated, hierarchical mergers.
The heavy remnants of BBH mergers are generically rapidly rotating, with spins centered around $\chi\sim0.7$.
If these ``second-generation'' black holes subsequently undergo additional mergers with ``first-generation'' black holes, the result will be a population of BBHs observed with very unequal mass ratios and preferentially large spins~\citep{Fishbach2017_hierarchical,Gerosa2017,Doctor2019,Rodriguez2019,Kimball2020,Gerosa2021}.
In the absence of any preferred directions, though, the result will be a \textit{broadening} of the $\chieff$ distribution towards smaller $q$ (e.g. negative $\beta$).
This prediction is at odds with the fact that the data prefer larger but \textit{preferentially positive} effective spins with smaller mass ratio.

As noted above, an alternative interpretation is that the BBHs observed in GWTC-2 are a mixture of populations arising from distinct formation channels~\citep[e.g.][]{franciolini2021,Wong2021,Zevin2021}; in this picture the measured $q$ vs. $\chieff$ anti-correlation arises from the presence of two or more sub-populations that manifest in the $\chieff-q$ plane.
As discussed in Sect.~\ref{sec:model-fit}, both GW190412 and GW190517 lie in the same area of this plane when reweighted to a population-informed prior, each exhibiting small $q$ and confidently positive $\chieff$.
\citet{Chia2021} have also recently argued that GW151226 also exhibits similar characteristics when re-analyzed with a waveform model including spin precessing and higher order radiation modes~\citep{Pratten2020}.
These events might together suggest the presence of a secondary sub-population appearing at low $q$ and high $\chieff$.

\begin{figure}
    \centering
    \includegraphics[width=0.43\textwidth]{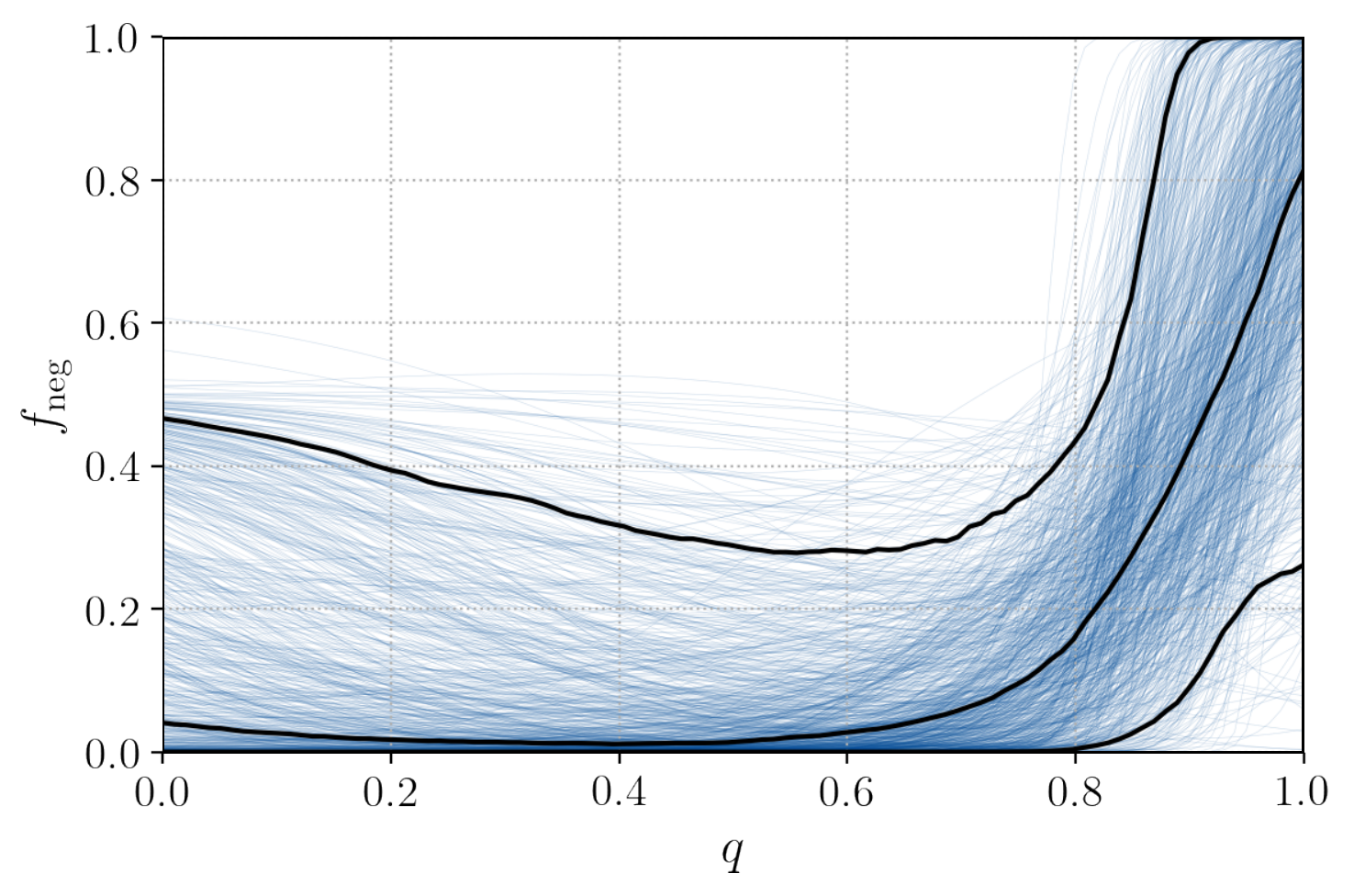}
    \caption{
    The inferred fraction of BBHs with negative effective spin as a function of mass ratio, assuming that the population is well-described by a Gaussian distribution.
    Blue traces mark individual draws from our posterior on population-level parameters (see Fig.~\ref{fig:corner-evolution}) and black traces mark the median and central 90\% credible bounds on $f_\mathrm{neg}$.
    When allowing for correlations between $q$ and $\chieff$, however, the data are fairly agnostic about the \textit{requirement} for negative effective spins; see Table~\ref{tab:evidences} and surrounding discussion.
    }
    \label{fig:f-neg}
\end{figure}

If we assume that GWTC-2 comprises a mixture of BBHs arising from isolated stellar evolution, with purely positive $\chieff$, and systems forming dynamically in stellar clusters, with isotropic spin orientations, we can attempt to characterize the implied mixture fraction between these two populations as a function of mass ratio.
Using our hierarchical measurement of $\{\mu_0,\sigma_0,\alpha,\beta\}$, in Fig.~\ref{fig:f-neg} we show our posterior on the fraction
    \begin{equation}
    f_\mathrm{neg}(q) = \int_{-1}^0 p(\chieff|q)\,d\chi
    \end{equation}
of events with negative effective spins versus $q$.
Individual blue traces show the result of single draws from our posterior on parameters governing the spin-mass ratio distribution, while solid black lines show the mean and central 90\% credible bounds on $f_\mathrm{neg}(q)$.
At $q\sim 1$, the data are consistent with half (or even more) of all systems exhibiting negative effective spins.
At low $q$, meanwhile, our inferred mean on the $\chieff$ distribution has shifted to larger positive values, and so the fraction of systems with negative $\chieff$ is likely quite small.

\begin{figure}
    \centering
    \includegraphics[width=0.43\textwidth]{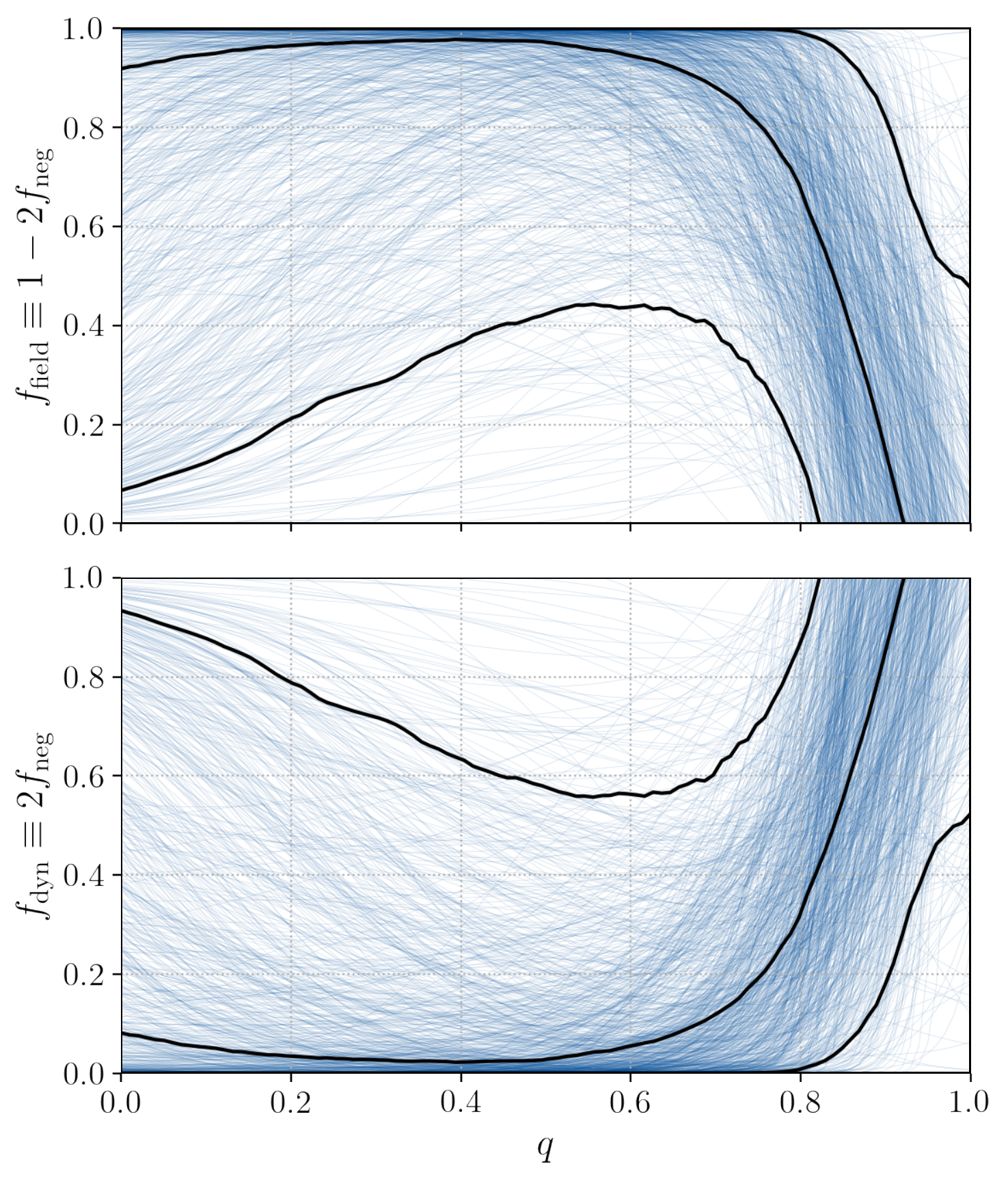}
    \caption{
    The implied fractions of BBHs arising from toy  ``dynamical'' (bottom) and ``isolated field binary'' (top) formation channels, assuming that the former yields a $\chieff$ distribution symmetric about zero, while the latter yields exclusively positive effective spins.
    Blue traces mark individual draws from the hyperposterior in Fig.~\ref{fig:corner-evolution}, and black traces denote medians and central 90\% credible bounds.
    Above $q\gtrsim 0.6$, the implied fraction of field events falls sharply with $q$, while at low mass ratios our uncertainties on $f_\mathrm{field}$ and $f_\mathrm{dyn}$ become large due to the small number of confidently unequal-mass BBHs.
    As in Fig.~\ref{fig:f-neg}, however, we note these results assume that the $\chieff$ distribution at a given $q$ is well-described by a Gaussian that extends to negative $\chieff$; this assumption is not necessarily independently supported by observation, as discussed below.
    }
    \label{fig:f-dyn-iso}
\end{figure}

If we very simplistically assume that BBHs formed dynamically in dense clusters exhibit a \textit{symmetric} $\chieff$ distribution, while BBHs arising in the field can possess only positive $\chieff$, then we obtain crude estimates
    \begin{equation}
    f_\mathrm{dyn}(q) = 2 f_\mathrm{neg}(q)
    \end{equation}
and
    \begin{equation}
    f_\mathrm{field}(q) = 1-2 f_\mathrm{neg}(q)
    \end{equation}
of the mixture fractions between channels as a function of $q$.
The result is shown in Fig.~\ref{fig:f-dyn-iso}.
At $q\sim 1$, the fraction of BBHs arising in clusters is at least $f_\mathrm{dyn} \approx 0.5$, or as high as one (note that some individual traces rise unphysically to values $f_\mathrm{dyn}>1$).
Below $q\approx 0.8$, meanwhile, it is likely  that $f_\mathrm{dyn}$ drops precipitously while the fraction $f_\mathrm{field}$ of systems formed in the field rises towards unity.
The rapid transition at $q\approx 0.8$ corresponds to the mass ratio at which $\mu_\chi(q)$ and $\sigma_\chi(q)$ are each most precisely measured; see Fig.~\ref{fig:traces}.
This result, with dynamically-formed systems exhibiting preferentially equal mass ratios while isolated BBHs possess unequal mass ratios, is surprising.
Theoretical modeling and population synthesis generally predict the opposite, such that BBHs formed in clusters have more unequal mass ratios than those formed in the field, due to the possibility of exchanges during three-body encounters and/or hierarchical mergers~\citep{Bouffanais2019,DiCarlo2020,Bouffanais2021}.

In constructing and interpreting Figs.~\ref{fig:f-neg} and \ref{fig:f-dyn-iso}, it is important to explore whether the constraints on $f_\mathrm{neg}(q)$ arise from informative data, or whether they are simply extrapolations based on the Gaussian model we adopt for $p(\chieff|q)$.
When neglecting the possibility of correlations between $q$ and $\chieff$, the \citet{O3a_pop} found evidence for negative effective spins among GWTC-2.
In particular, when fitting the $\chieff$ with a Gaussian truncated between $\chi_\mathrm{eff,min}\leq \chieff \leq 1$, for some unknown $\chi_\mathrm{eff,min}$, this lower truncation bound was inferred to be negative at high credibility.
This evidence for negative effective spins was corroborated via an alternative model that directly fit for the spin magnitudes and tilt angles of component black holes.
\citet{Callister2020-sn} hierarchically modeled the $\chieff$ distribution not as Gaussian, but as the convolution of an underlying component spin distribution with a distribution of natal kicks experienced during the core-collapse of black hole progenitors; the results of this analysis too indicated the presence of negative $\chieff$.
More recently, though, \citet{Roulet2021} have argued that present detections \textit{lack} strong evidence for the existence of negative $\chi_\mathrm{eff}$ and/or extreme spin-orbit misalignment; see further discussion below.

\begin{table}
\begin{center}
\setlength{\tabcolsep}{12pt}
\begin{tabular}{ l | r }
\hline
\hline
Model & $\ln\,\mathrm{Evidence}$ \\
\hline
Correlated $\chieff$ \& $q$, & $\evidenceYesEvolYesNeg\,(\pm \evidenceYesEvolYesNegError)$ \\
Correlated $\chieff$ \& $q$, with $\chieff\geq 0$ & $\evidenceYesEvolNoNeg\,(\pm \evidenceYesEvolNoNegError)$ \\
Independent $\chieff$ \& $q$, & $\evidenceNoEvolYesNeg\,(\pm \evidenceNoEvolYesNegError)$ \\
Independent $\chieff$ \& $q$, with $\chieff\geq 0$ & $\evidenceNoEvolNoNeg\,(\pm \evidenceNoEvolNoNegError)$ \\
\hline
\hline
\end{tabular}
\caption{
Relative Bayesian evidences for the four model variants discussed in Sects.~\ref{sec:residuals}, \ref{sec:model-fit}, and \ref{sec:implications}, encompassing the possibilities that $q$ and $\chieff$ are [Eq.~\eqref{eq:pchi-evol}] or are not [Eq.~\eqref{eq:pchi}] correlated, together with the existence or non-existence of BBHs with negative effective spins.
Evidences are obtained using \textsc{Dynesty}~\citep{Speagle2020}, and the stated uncertainties are obtained by computing each evidence ten times and recording the standard deviation among runs.
We choose a normalization such the most disfavored model (uncorrelated parameters with purely positive $\chieff$) has a mean log-evidence of zero.
}
\label{tab:evidences}
\end{center}
\end{table}

These conclusions, though, were generally based on models that did not include correlations between $q$ and $\chieff$.
To understand the extent to which negative effective spins remain required by the data, we use the \textsc{Dynesty} nested sampler~\citep{Speagle2020} to compute Bayesian evidences for two additional models beyond those introduced in Sects.~\ref{sec:residuals} and \ref{sec:model-fit} above:
    \begin{itemize}
        \item Independent $\chieff$ and $q$, with purely positive $\chieff$.
        In this case, Eq.~\eqref{eq:pchi} is truncated on $[0,1]$, rather than $[-1,1]$, with a restricted prior $0\leq \mu \leq 1$ on its mean.
        \item Correlated $\chieff$ and $q$, with purely positive $\chieff$.
        Equation~\eqref{eq:pchi-evol} is truncated on $[0,1]$, and we adopt a restricted prior enforcing $0\leq \mu_{\chi,0} \leq 1$.
    \end{itemize}
In each case we use the same modeling assumptions and priors as listed in Appendix~\ref{sec:hierachical-appendix}, unless otherwise noted.

Table~\ref{tab:evidences} lists the (natural) log-evidences for these two ``purely-positive spin'' models, in addition to the evidences for the models of Sects.~\ref{sec:residuals} and \ref{sec:model-fit} and discussed earlier.
The log-evidences are scaled such that the most-disfavored model has $\ln\,\mathrm{Evidence} = 0$.
The most favored model is one that allows for both a $\chieff-q$ correlation \textit{and} negative effective spins, with a Bayes factor $\ln\mathcal{B} = \evidenceYesEvolYesNeg$ relative to a model including neither effect.
However, allowing for $\chieff-q$ correlations \textit{weakens} evidence for negative effective spins;
we find a nearly-uninformative Bayes factor $\ln\mathcal{B} = \bayesNegWhenCorrelated$ between our most favored model and one that allows for correlated $\chieff$ and $q$ but \textit{without} negative $\chieff$.
As the Advanced LIGO and Virgo detectors continue adding to the list of BBH detections, it will be important to revisit this model comparison, refining the evidence (or lack thereof) for negative effective spins and understanding the extent to which conclusions regarding negative $\chieff$ and/or correlations between parameters can confound one another.

\begin{figure}
    \centering
    \includegraphics[width=0.48\textwidth]{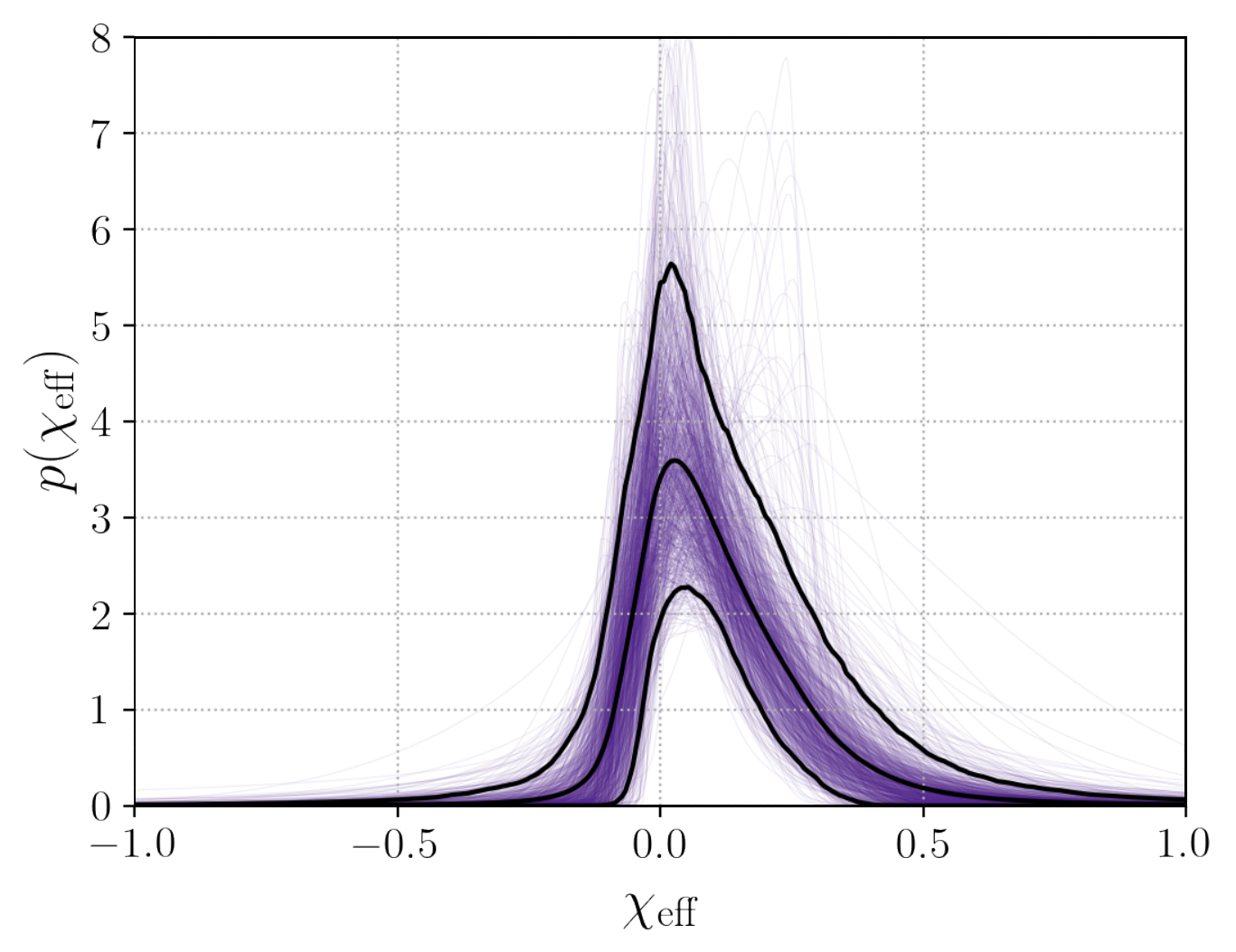}
    \caption{
    The marginal $\chi_\mathrm{eff}$ distribution obtained by integrating our inferred joint distribution $p(q,\chi_\mathrm{eff})$ over $q$.
    Blue traces give single draws from our hyperposterior in Fig.~\ref{fig:corner-evolution}, while black traces denote the median and central 90\% credible bounds.
    The marginal distribution exhibits a sharp peak about zero, with an extended tail towards large and positive $\chi_\mathrm{eff}$.
    The peak occurs at $q\approx1$, where the $\chi_\mathrm{eff}$ grows narrow, while the extended tail arises from the shift in $\chi_\mathrm{eff}$ towards more positive values with lower $q$.
    }
    \label{fig:marginal-chi}
\end{figure}

We note that, in gauging the evidence for/against negative $\chi_\mathrm{eff}$, we are still assuming that BBHs are well-described by a \textit{single} population in the $q-\chi_\mathrm{eff}$ plane.
\citet{Roulet2021} have recently demonstrated that conclusions regarding the presence of negative $\chi_\mathrm{eff}$ depend strongly on whether or not one allows for a \textit{second} population of events with near-vanishing spins.
When hierarchically measuring the BBH $\chi_\mathrm{eff}$ distribution using a mixture model that included such a vanishing spin sub-population, they argued that evidence for negative effective spins is significantly reduced.
It is interesting to note that a $\chi_\mathrm{eff}$ distribution with a vanishing-spin sub-population arises naturally from our analysis here.
Figure~\ref{fig:marginal-chi} shows the marginal $\chi_\mathrm{eff}$ distribution implied by our results, having integrated the joint mass ratio-spin distribution $p(q,\chi_\mathrm{eff})$ over $q$.
The result is an asymmetric distribution consistent with the marginal $\chi_\mathrm{eff}$ model favored by \citet{Roulet2021}, exhibiting both a peak at $\chi_\mathrm{eff} \approx 0$ (occurring at high $q$) and a shoulder extending towards positive values (arising from the low-$q$ population).

Finally, it should be emphasized that despite our focus on BBH formation via ``canonical'' field binary and cluster scenarios in this discussion, these are certainly not the only two  options that exist.
Other possible avenues for BBH formation and merger include hierarchical triples~\citep{Liu2017,Liu2018,Antonini2018,Rodriguez_triple_2018,Martinez2020} and BBH assembly in the disks of active galactic nuclei~\citep[AGN;][]{Mckernan2018,Mckernan2019,Stone2017,Tagawa2020}; it is not well-understood what kinds of $q$ and $\chieff$ correlations might arise under these alternative scenarios.
AGN disks, in particular, offer opportunities for hierarchical BBH mergers while maintaining a preferred direction (set by the accretion disk's angular momentum), and thus might be able to explain the $q-\chieff$ anti-correlation we observe here.

\section{Conclusions}

In this paper, we have presented evidence for an intrinsic anti-correlation between the effective spins and mass ratios of detected BBH mergers.
Under a hierarchical analysis of the BBH population using Advanced LIGO and Virgo's GWTC-2 catalog, we find that equal-mass binaries preferentially exhibit a narrow $\chieff$ distribution centered at zero, while unequal mass events favor a $\chieff$ distribution systematically shifted towards larger, positive values.

This anti-correlation is unexpected.
As discussed in Sect.~\ref{sec:implications}, it is not clear what physical processes might give rise to such an effect; in fact, standard ideas about BBH formation via isolated stellar evolution and dynamical assembly in dense clusters make predictions at odds with the trend we discover here.
For this reason, we have attempted to critically examine and test various sources of bias or systematic uncertainty that might lead to a spurious conclusions regarding the joint $\chieff-q$ distribution in BBH mergers.
In Sect.~\ref{sec:tests}, we found that the observed anti-correlation between $q$ and $\chieff$ could not be clearly attributed to selection effects, measurement degeneracies, or a small number of outlier events among the BBH population.

Looking ahead, the continued detection of additional BBHs by Advanced LIGO and Virgo \citep[to be joined soon by the KAGRA experiment;][]{kagra} will be crucial to better understand the intrinsic relationship between $q$ and $\chieff$, further bolstering our conclusions or pointing instead to a statistical fluctuation or as-of-yet unknown systematic.
Simultaneously, it will be valuable to re-evaluate our understanding of BBH formation in light of the results presented here, exploring which formation channel (or combination of channels) can accommodate the observed trend in $q$ and $\chieff$.
If confirmed to be astrophysical, the anti-correlated mass ratios and effective spins of BBHs may offer a firm observational foothold on our path towards understanding the origin of black hole mergers.

{
\vspace{3mm}
\noindent\textit{Acknowledgements}.
We thank Katerina Chatziioannou for helpful conversation, and our anonymous referees whose suggestions have greatly improved this manuscript.
The Flatiron Institute is a division of the Simons Foundation, supported through the generosity of Marilyn and Jim Simons.
This material is based upon work supported by NSF’s LIGO Laboratory which is a major facility fully funded by the National Science Foundation.
We are grateful for computational resources provided by the LIGO Laboratory and supported by National Science Foundation Grants PHY-0757058 and PHY-0823459.
This research has made use of data, software and/or web tools obtained from the Gravitational Wave Open Science Center (\url{https://www.gw-openscience.org/}), a service of LIGO Laboratory, the LIGO Scientific Collaboration and the Virgo Collaboration. LIGO Laboratory and Advanced LIGO are funded by the United States National Science Foundation (NSF) as well as the Science and Technology Facilities Council (STFC) of the United Kingdom, the Max-Planck-Society (MPS), and the State of Niedersachsen/Germany for support of the construction of Advanced LIGO and construction and operation of the GEO600 detector. Additional support for Advanced LIGO was provided by the Australian Research Council. Virgo is funded, through the European Gravitational Observatory (EGO), by the French Centre National de Recherche Scientifique (CNRS), the Italian Istituto Nazionale di Fisica Nucleare (INFN) and the Dutch Nikhef, with contributions by institutions from Belgium, Germany, Greece, Hungary, Ireland, Japan, Monaco, Poland, Portugal, and Spain.
}

{
\vspace{3mm}
\noindent\textit{Data \& code availability}.
The code used to produce the results in this paper and the resulting data products are available at \url{https://github.com/tcallister/BBH-spin-q-correlations}.
}

{
\vspace{3mm}
\noindent\textit{Software used}.
\textsc{astropy}~\citep{astropy1,astropy2},
\textsc{bilby}~\citep{Ashton2019,Romero-Shaw2020},
\textsc{dynesty}~\citep{Speagle2020},
\textsc{emcee}~\citep{emcee},
\textsc{h5py}~\citep{h5py},
\textsc{Matplotlib}~\citep{matplotlib},
\textsc{NumPy}~\citep{numpy},
\textsc{PyCBC}~\citep{pycbc},
\textsc{pandas}~\citep{pandas1,pandas2},
\textsc{SciPy}~\citep{scipy}
}

%TC:ignore
\appendix
\section{Hierarchical inference of the BBH population}
\label{sec:hierachical-appendix}

Here we discuss details of our hierarchical inference of the BBH population with  GWTC-2.
We model the primary mass distribution following the \textsc{Power Law + Peak} model of \citet{Talbot2018} and \citet{O3a_pop}, in which primary BBH masses are described as a mixture
    \begin{equation}
    p(m_1|f_p,\lambda,\mu_m,\sigma_m,m_\mathrm{max}) = f_p\,P(m_1|\lambda,m_\mathrm{max}) + (1-f_p) N(m_1|\mu_m,\sigma_m,m_\mathrm{max})
    \label{eq:pm}
    \end{equation}
between a power law $P(m_1|\lambda,m_\mathrm{max}) \propto m_1^\lambda$ and a Gaussian $N(m_1|\mu_m,\sigma_m,m_\mathrm{max}) \propto \exp\left[-\frac{(m_1-\mu_m)^2}{2\sigma_m^2}\right]$, normalized across the range $5\,M_\odot \leq m_1 \leq m_\mathrm{max}$.
We assume a redshift distribution that is proportional to the differential comoving volume $\frac{dV_c}{dz}$, with a possible evolution in the merger rate towards higher redshift~\citep{Fishbach2018,Callister2020},
    \begin{equation}
    p(z|\kappa) \propto \frac{1}{1+z} \frac{dV_c}{dz} \left(1+z\right)^\kappa.
    \label{eq:pz}
    \end{equation}
The additional factor of $(1+z)^{-1}$ in Eq.~\eqref{eq:pz} converts a uniform-in-time source-frame distribution to our detector frame.
We adopt cosmological parameters consistent with those reported in \citet{Planck2016}.

As discussed in the main text, we explore two related models for the mass ratio and effective spin distribution of BBHs.
In Sect.~\ref{sec:residuals}, we first consider the case in which these parameters are uncorrelated, describing the conditional distribution $p(q|m_1,\gamma)$ of mass ratios as a power law with index $\gamma$ and the distribution $p(\chieff|\mu_\chi,\sigma_\chi)$ of effective spin parameters as a Gaussian with mean $\mu_\chi$ and standard deviation $\sigma_\chi$; see Eqs.~\eqref{eq:pq} and \eqref{eq:pchi}.
In Sect.~\ref{sec:model-fit} and beyond, we subsequently expand this model to allow for population-level correlations between the mass ratios and spins of BBHs.
In particular, we preserve the Gaussian form for $p(\chieff)$, but now allow its mean and standard deviation to vary as a function of mass ratio; see Eqs.~\eqref{eq:pchi-evol}, \eqref{eq:mean-chi}, and \eqref{eq:sig-chi}.

Together, we hierarchically fit the collection of hyperparameters that govern the BBH population by considering the $N_{\rm det} = 44$ BBH candidates among GWTC-2 with false alarm rates below one per year~\citep{gwtc2}.
We make use of the parameter estimation results described in \citet{gwtc1,gwtc2} and made publicly available through the Gravitational-Wave Open Science Center~\citep{losc,data_release1,data_release,gwosc,open-data}.
For BBHs first announced in GWTC-1, we use the \texttt{Overall\_Posterior} samples formed by the union of results under two different waveform families.
For BBHs observed in LIGO and Virgo's O3a observing run, we use the \texttt{PrecessingSpinIMRHM} samples generated by waveforms including the effects of spin precession and higher-order modes.

Given posteriors $p(\theta_i|d_i)$ on the individual parameters $\theta_i$ (e.g. component masses, spins, etc.) of each event conditioned on its observed data $d_i$, the corresponding posterior on the population parameters $\Lambda$ is~\citep{Loredo2004,Taylor2018,Mandel2019,Vitale2020}
    \begin{equation}
    p(\Lambda \,|\, \{d_i\})
        \propto p(\Lambda)\, \xi^{-N_{\rm det}}(\Lambda)
        \prod_{i=1}^{N_{\rm det}}
        \int d\theta_i\, p(\theta_i|d_i)
        \frac{p(\theta_i|\Lambda)}{p_\mathrm{pe}(\theta_i)}.
    \label{eq:likelihood-integral}
    \end{equation}
This form of the likelihood includes implicit marginalization over the overall \textit{rate} of BBH mergers, using a log-uniform prior on the expected number of detections~\citep{Fishbach2018,Mandel2019}.
In Eq.~\eqref{eq:likelihood-integral}, $p_\mathrm{pe}(\theta_i)$ is the default prior adopted for purposes of parameter estimation and  $\xi(\Lambda)$ is the population-weighted detection efficiency, discussed further below.
Parameter estimation for GWTC-2 is achieved using priors that are uniform in \textit{detector} frame masses and Euclidean volume, corresponding to an implicit prior~\citep{gwtc1,gwtc2,Callister2021}
    \begin{equation}
    p_\mathrm{pe}(m_1,m_2,z) \propto (1+z)^2 D_L^2(z) \frac{dD_L}{dz}
    \label{eq:prior-m-z}
    \end{equation}
on source-frame masses and redshift, where $D_L(z)$ is the luminosity distance at redshift $z$.
Component spin priors are uniform in magnitude and isotropic in orientation, corresponding to an effective spin prior $p_\mathrm{pe}(\chieff)$ given by Eq.~(10) of \citet{Callister2021}.
Meanwhile, $p(\Lambda)$ is our prior on the population-level parameters. Unless stated otherwise, we use the priors listed in Table~\ref{tab:priors}.
We still cannot yet employ Eq.~\eqref{eq:likelihood-integral}, since we do not have direct access underlying posteriors $p(\theta_i\,|\,d_i)$, but instead have a collection of discrete \textit{samples} drawn from each event's posterior.
Instead, we replace integration over $p(\theta_i|d_i)$ with  an ensemble average taken over the posterior samples associated with each event:
    \begin{equation}
    p(\Lambda \,|\, \{d_i\})
        \propto p(\Lambda)\,\xi^{-N_{\rm det}}(\Lambda)
        \prod_{i=1}^{N_{\rm det}}
        \bigg\langle
        \frac{p(\theta_i|\Lambda)}{p_\mathrm{pe}(\theta_i)}.
        \bigg\rangle
    \label{eq:likelihood-sum}
    \end{equation}
We sample over Eq.~\eqref{eq:likelihood-sum} using the \textsc{emcee} Markov Chain Monte Carlo sampler~\citep{emcee}, to obtain the posteriors shown in Figs.~\ref{fig:corner-evolution}, \ref{fig:default-corner}, and \ref{fig:other-params}.
When computing and comparing Bayesian evidences in Sect.~\ref{sec:implications}, we instead implement and integrate over Eq.~\eqref{eq:likelihood-sum} using the \textsc{Dynesty} nested sampler~\citep{Speagle2020}.
    
\begin{table}
\begin{center}
 \renewcommand{\arraystretch}{1.05}
 \begin{tabular}{c l l}
 \hline
 \hline
 Parameter & Prior & Defined in \\
 \hline
$\mu_{\chi,0}$ & $U(-1,1)$ & Eqs.~\eqref{eq:pchi} \& \eqref{eq:pchi-evol} \\
$\log_{10}\sigma_{\chi,0}$ & $U(-1.5,0.5)$ & Eqs.~\eqref{eq:pchi} \& \eqref{eq:pchi-evol}  \\
$\alpha$ & $U(-2.5,1)$ & Eq.~\eqref{eq:pchi-evol}\\
$\beta$ & $U(-2,1.5)$ & Eq.~\eqref{eq:pchi-evol} \\
$m_\mu$ & $U(20\,M_\odot,100\,M_\odot)$ & Eq.~\eqref{eq:pm} \\
$m_\sigma$ & $U(1\,M_\odot,10\,M_\odot)$ & Eq.~\eqref{eq:pm} \\
$f_p$ & $U(0,1)$ & Eq.~\eqref{eq:pm} \\
$\lambda$ & $U(-5,4)$ & Eq.~\eqref{eq:pm} \\
$\gamma$ & $U(-2,10)$ & Eq.~\eqref{eq:pq} \\
$ m_{\rm max}$ & $U(60\,M_\odot,100\,M_\odot)$ & Eq.~\eqref{eq:pm} \\
$\kappa$ & $N(0,6)$ & Eq.~\eqref{eq:pz} \\
 \hline
 \hline
\end{tabular}
\caption{Priors adopted for the hyperparameters with which we describe the mass, spin, and redshift distributions of BBHs.
Here, $U(\mathrm{min},\mathrm{max})$ denotes a uniform prior distribution between the given minimum and maximum values, while $N(a,b)$ is a normal distribution with mean $a$ and standard deviation $b$.
When modeling the $\chieff$ distribution in Sect.~\ref{sec:residuals} \textit{without} any correlations with $q$, our priors on $\mu_\chi$ and $\sigma_\chi$ are identical to those listed here for $\mu_{\chi,0}$ and $\mu_{\chi,0}$.
}
\label{tab:priors}
\end{center}
\end{table}

In Eqs.~\eqref{eq:likelihood-integral} and \eqref{eq:likelihood-sum}, the detection efficiency $\xi(\Lambda)$ quantifies the fraction of events that we expect to pass our detection criteria, given a population described by $\Lambda$:
    \begin{equation}
    \xi(\Lambda) = \int d\theta \, P_{\rm det}(\theta)\, p(\theta|\Lambda).
    \end{equation}
Here, $P_\mathrm{det}(\theta)$ is the detection probability for a particular event with parameters $\theta$.
The detection efficiency corrects for search selection effects and so is critical to accurately calculate.
We estimate $\xi(\Lambda)$ using the injection campaign reported in \citet{O3a_pop,injections},
selecting successfully found injections (with recovered false alarm rates below one per year in at least one pipeline) and reweighting to the proposed population $\Lambda$:
    \begin{equation}
    \xi(\Lambda) = \frac{1}{N_\mathrm{inj}} \bigg\langle
        \frac{p(\theta|\Lambda)}{p_{\rm inj}(\theta)}
        \bigg\rangle_{\rm Found\, injections},
    \label{eq:sel-effects}
    \end{equation}
where $N_\mathrm{inj}$ is the total number of injections (including those that are \textit{not} recovered) and $p_{\rm inj}(\theta)$ is the reference distribution from which injections were drawn.
As discussed in \citet{injections}, the injected masses follow $p_\mathrm{inj}(m_1) \propto m_1^{-2.35}$ for $2\,M_\odot \leq m_1 \leq 100\,M_\odot$ and $p(q|m_1) \propto q^2$.
The injections additionally have purely aligned component spins ($\theta_1 = \theta_2 = 0$ or $\pi$) distributed uniformly between $-1 \leq \chi_{z} \leq 1$; the corresponding distribution of $\chieff$ is~\citep{Callister2021}
    \begin{equation}
    p_\mathrm{inj}(\chieff|q) = \begin{cases}
        \dfrac{(1+q)^2(1+\chieff)}{4q}
            & \left(\chieff\geq -1 \quad \& \quad \chieff<-\dfrac{1-q}{1+q}\right) \\
        \dfrac{1+q}{2}
            & \left(\chieff\geq-\dfrac{1-q}{1+q} \quad\&\quad \chieff\leq\dfrac{1-q}{1+q}\right) \\
        \dfrac{(1+q)^2(1-\chieff)}{4q}
            & \left(\chieff>\dfrac{1-q}{1+q} \quad \& \quad \chieff\leq 1\right)
        \end{cases}
    \label{eq:prior-inj-chieff}
    \end{equation}
    
One operation frequently performed in this paper is the reweighting of single-event posteriors from the default prior $p_\mathrm{pe}(\theta)$ adopted for parameter estimation to some new \textit{population-informed} prior based on our hierarchical inference of $\Lambda$.
If we denote by $p_\mathrm{pe}(\theta|d)$ the posterior on a particular event obtained through standard parameter estimation, then via Bayes' theorem, the reweighted posterior corresponding to some \textit{fixed} value of $\Lambda$ is
    \begin{equation}
    p(\theta | d,\Lambda)
        \propto p_\mathrm{pe}(\theta|d)
            \left(\frac{p(\theta|\Lambda)}{p_\mathrm{pe}(\theta)}\right).
    \label{eq:reweighting}
    \end{equation}
We do not, of course, measure $\Lambda$ perfectly.
What we instead want is a reweighted posterior that has been marginalized over our uncertainty on the nature of the underlying population.
Given an original set of samples $\{\theta\}_\mathrm{pe}$ drawn from $p_\mathrm{pe}(\theta|d)$ and a set of hyperparameter samples $\{\Lambda\}$ generated by our population fit, we can obtain a new set of samples drawn from this reweighted and marginalized posterior by doing the following:
    \begin{enumerate}
        \item Randomly select a hyperparameter sample $\Lambda_i \in\{\Lambda\}$
        \item For each sample $\theta_j \in \{\theta\}_{\mathrm{pe}}$, compute the weights $w_j =  p(\theta_j|d,\Lambda_i)/p_\mathrm{pe}(\theta_j)$ that correspond to the parenthetical in Eq.~\eqref{eq:reweighting}
        \item Randomly select and store a posterior sample $\theta_j$ according to weights $w_j$
        \item Repeat
    \end{enumerate}
As is discussed in \citet{Callister-reweighting} and \citet{Farr-reweighting}, this algorithm can be shown to properly avoid the ``double-counting'' of information; the updated prior imposed on any one event is itself informed only by the \textit{other} events in one's sample, excluding the particular event of interest.

Another operation performed in Sects.~\ref{sec:residuals} and \ref{sec:model-fit} is the generation of \textit{predicted} observations, marginalized over the parameters of a particular population model.
This proceeds in a manner similar to posterior reweighting, but instead using the pipeline injection sets that inform our calculation of $\xi(\Lambda)$ above:
    \begin{enumerate}
        \item Randomly select a hyperparameter sample $\Lambda_i \in\{\Lambda\}$
        \item For each found injection, with parameters $\theta_j$, compute weights $w_j =  p(\theta_j|\Lambda_i)/p_\mathrm{inj}(\theta_j)$ to reweight from the injected reference distribution to the proposed population
        \item Randomly select and store one found injection according to the weights $w_j$
        \item Repeat
    \end{enumerate}
    
\section{Which events drive the spin-mass ratio correlation?}
\label{sec:most-important-events}

In Sect.~\ref{sec:tests}, we verified that our measurement of an anti-correlation between $q$ and $\chi_\mathrm{eff}$ is not driven solely by GW190517 and GW190412.
Here, we try to understand more generally \textit{which} events are driving (or resisting) our result.
For every event in our sample we compute a Bayes factor between two fixed BBH population models: a fixed ``correlated'' population consistent with our posteriors on the expanded model of Sect.~\ref{sec:model-fit}, and a fixed ``uncorrelated'' population consistent with results from our initial model in Sect.~\ref{sec:residuals}.
The events with Bayes factors favoring the correlated population are likely those same events driving our measurement of $\alpha<0$, while events that favor the uncorrelated population likely act to resist a $q-\chi_\mathrm{eff}$ anti-correlation.
For both models, we choose hyperparameters consistent with the median values inferred by a complete hierarchical analysis.
Specifically, we take $m_\mu = 33\,M_\odot$, $m_\sigma = 5\,M_\odot$, $f_p = 0.1$, $\lambda = -2.5$, $\gamma = 1.1$, $m_\mathrm{min} = 5\,M_\odot$, $m_\mathrm{max} = 100\,M_\odot$, and $\kappa = 2.7$.
For the ``correlated'' population, we take $\mu_{\chi,0} = 0.2$, $\log_{10}\sigma_{\chi,0} = -1$, $\alpha = -0.45$, and $\beta = 0$, while for the ``uncorrelated'' population we fix $\mu_\chi = 0.05$ and $\log_{10}\sigma_\chi = -1$.

\begin{figure*}[t!]
    \centering
    \includegraphics[width=0.95\textwidth]{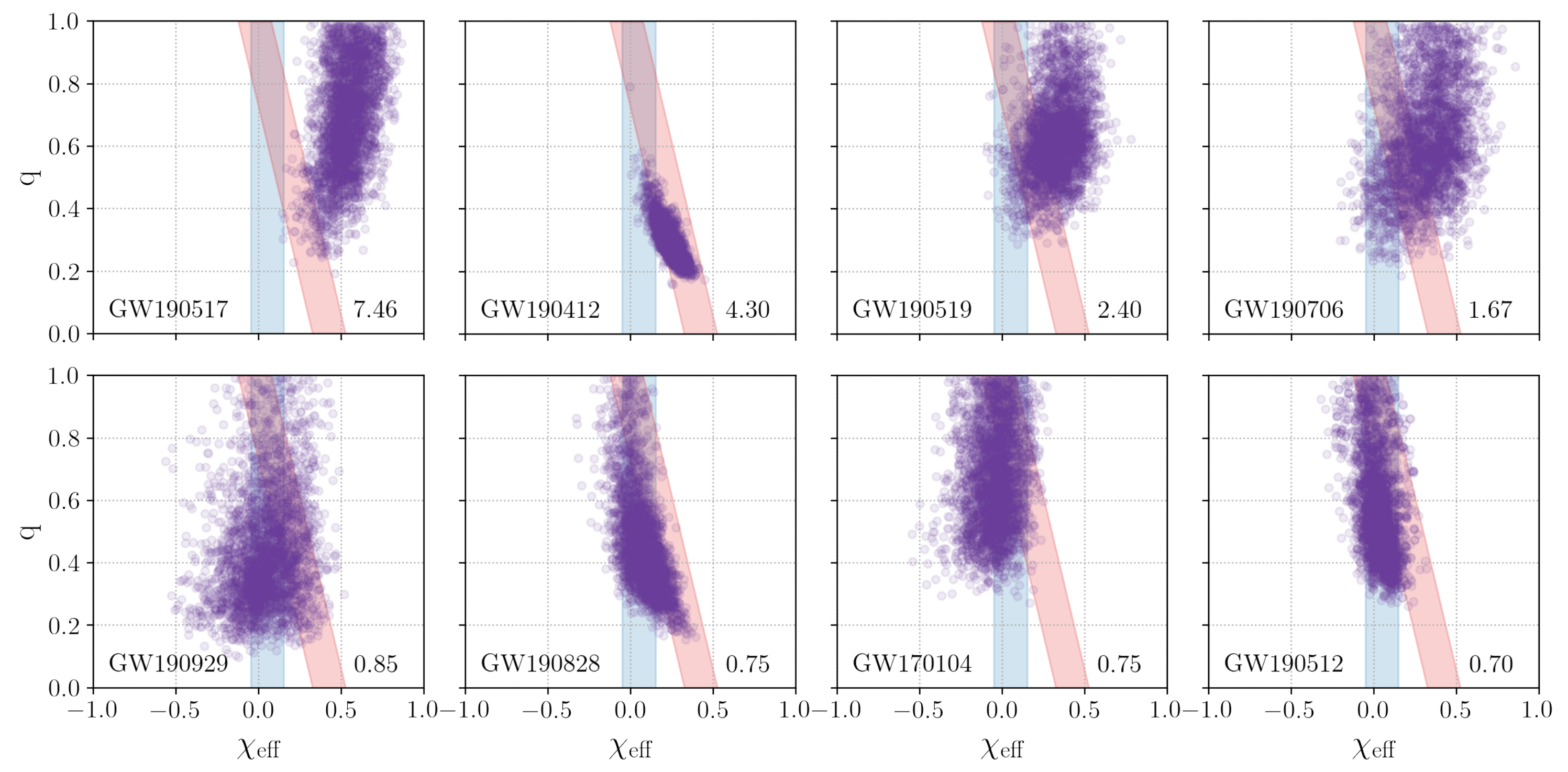}
    \caption{
    The events with the largest (top row) and smallest (bottom row) Bayes factors, indicated in the lower right corner of each panel, between population models with and without correlations between $q$ and $\chi_\mathrm{eff}$ as described in Appendix~\ref{sec:most-important-events}.
    In each plot, blue points mark posterior samples for each event under default parameter estimation priors, while the red and blue bands show one-sigma bounds $\mu_\chi(q)\pm \sigma_\chi(q)$ on the $\chi_\mathrm{eff}$ distribution under the correlated and uncorrelated populations.
    In Sect.~\ref{sec:tests}, we worried that large measurement degeneracies between $q$ and $\chi_\mathrm{eff}$ might spuriously cause an apparent correlation between these parameters among the BBH population.
    Of the events shown here that most favor a population-level correlation, though, none exhibit large curving measurement degeneracies. 
    }
    \label{fig:most-important}
\end{figure*}

In Fig.~\ref{fig:most-important}, we show the four events with the largest (top row) and smallest (bottom row) Bayes factors between our fixed correlated and uncorrelated population models.
Blue points show each events' posterior samples (under default parameter estimation priors) and the correlated vs. uncorrelated Bayes factor is printed in the lower right corner of each subplot.
For reference, the filled blue band shows the range of spins $\mu_\chi \pm \sigma_\chi$ primarily supported by the uncorrelated population, while the sloped red band shows spin range $\mu_\chi(q)\pm \sigma_\chi(q)$ favored in the correlated population.

As anticipated, GW190517 and GW190412 most strongly identify as members of the correlated population.
Several other events with posterior morphologies similar to that of GW190517, however, are also picked out as favoring the correlated population model.
Most notably, \textit{none of the events that most favor a $q-\chi_\mathrm{eff}$ anti-correlation exhibit the long, curving degeneracies which we discussed in Sect.~\ref{sec:tests}.}
This further bolsters our confidence that the $q-\chi_\mathrm{eff}$ anti-correlation identified among the BBH population is not due to the combined ``leakage'' of degenerate measurements for individual events.
The common morphologies among GW190517, GW190519, and GW190706 additionally offer intuition as to why excluding GW190412 from our analysis yields an even \textit{stronger} anti-correlation; see Fig.~\ref{fig:out-in}.
If we were to ignore GW190412 and redraw a red band passing through the posteriors of GW190517, GW190519, and GW190706, our band would likely be far shallower than that currently plotted in Fig.~\ref{fig:most-important}, corresponding to a more negative value of $\alpha$.

Meanwhile, the four events with the strongest Bayes factors \textit{against} a spin-mass ratio correlation display a variety of morphologies.
GW170104 favors low spins and a mass ratio near unity, while GW190929 and GW190512 favor mass more unequal mass ratios.
Interestingly, GW190828 displays the same anti-correlated \textit{measurements} that we considered in Sect.~\ref{sec:tests}.
However, this event resists our measurement of negative $\alpha$, rather than driving it.

\section{Detailed Parameter Estimation Results}
\label{sec:pe-appendix}

\begin{figure*}[t!]
    \centering
    \includegraphics[width=0.6\textwidth]{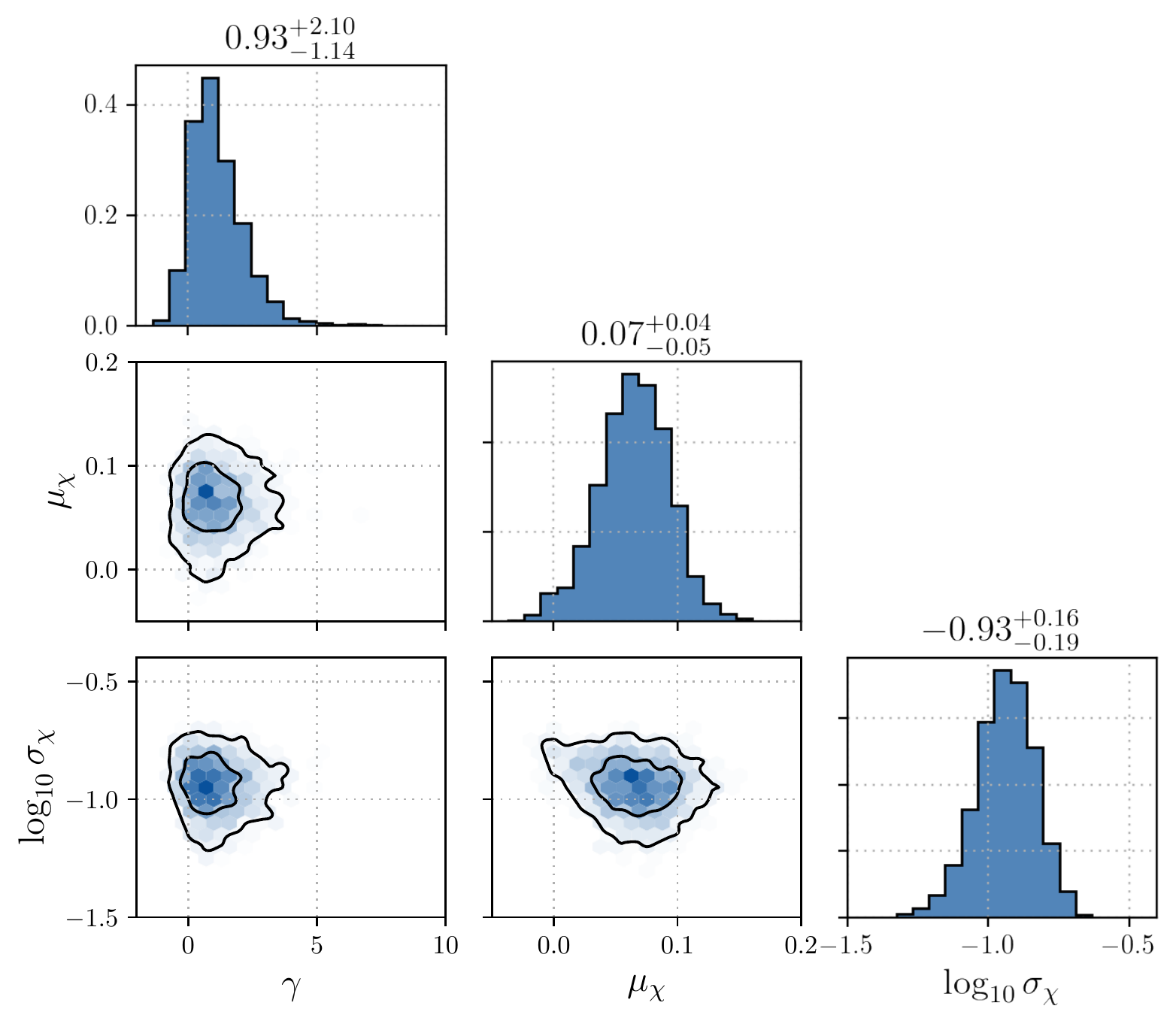}
    \caption{
    Posteriors on the parameters governing the mass ratio and effective spin distributions of BBHs, under the base model in Sect.~\ref{sec:residuals} with no correlations between $q$ and $\chieff$.
    The labels above each one-dimensional posterior give the medians and central 90\% credible uncertainties on each parameter, while the contours in each two-dimensional posterior enclose the 50\% and 90\% credible regions.
    Posteriors on the remaining parameters governing the BBH primary mass and redshift distributions are effectively identical to those recovered in \citet{O3a_pop}.
    For reference, though, the one-dimensional marginalized posteriors on these other parameters are shown as unfilled grey histograms in Fig.~\ref{fig:other-params} below.
    }
    \label{fig:default-corner}
\end{figure*}

\begin{figure*}[t!]
    \centering
    \includegraphics[width=0.8\textwidth]{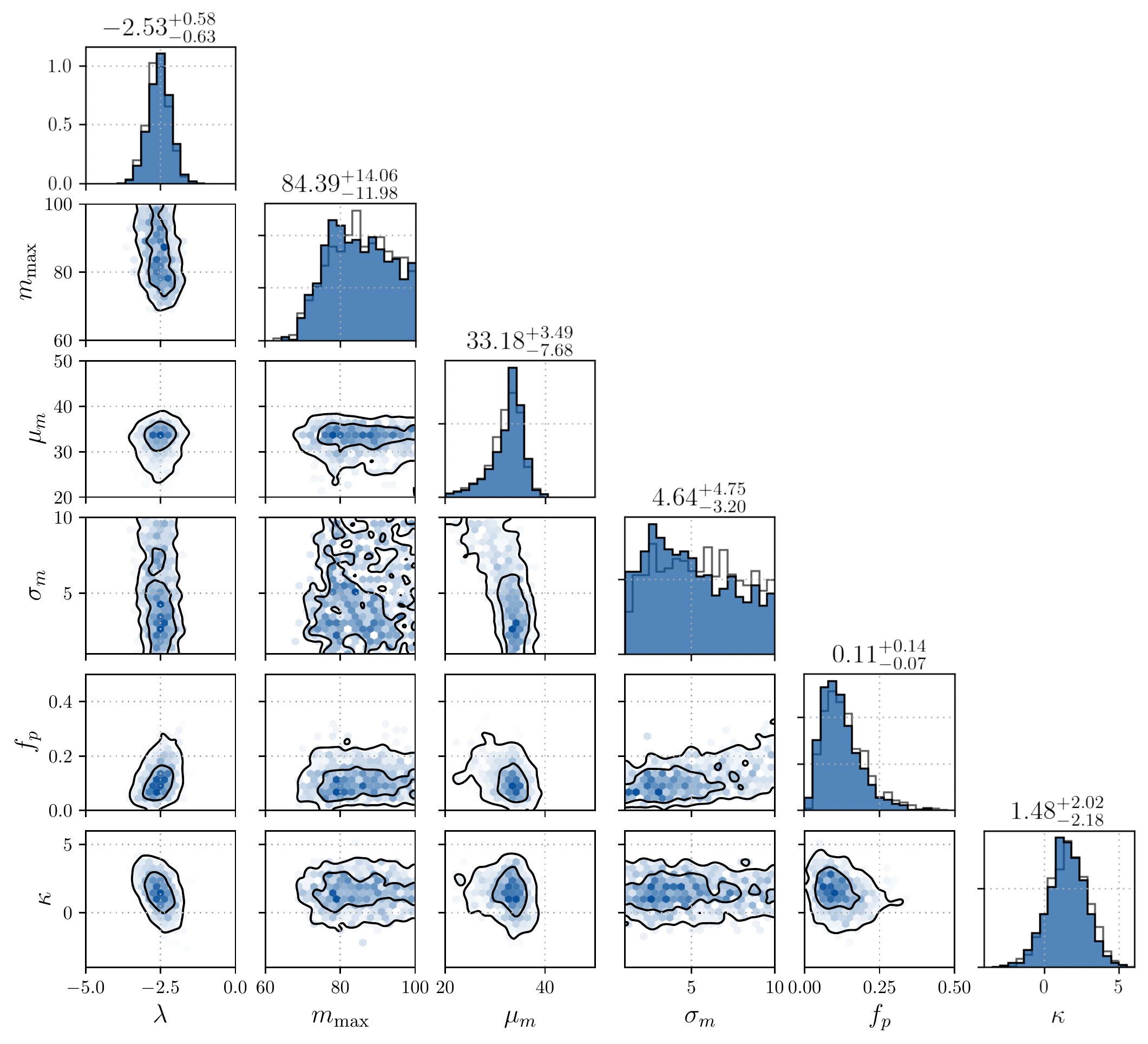}
    \caption{
    Posteriors on parameters governing the primary mass and redshift distributions of BBHs, under the expanded model described in Sect.~\ref{sec:model-fit} that allows for correlations between $q$ and $\chieff$.
    Posteriors on the parameters governing the $q$ and $\chieff$ distributions are shown in Fig.~\ref{fig:corner-evolution} in the main text.
    The numbers above each one-dimensional posterior quote medians and central 90\% credible uncertainties, and the two contours appearing in each two-dimensional posterior enclose the central 50\% and 90\% credible regions.
    For comparison, the unfilled grey histograms show the constraints on these parameters obtained under the base model of Sect.~\ref{sec:residuals} that \textit{neglects} any correlations between $q$ and $\chieff$; see also Fig.~\ref{fig:default-corner} above.
    }
    \label{fig:other-params}
\end{figure*}

Here, we show additional parameter estimation results neglected in the main text.
Figure~\ref{fig:default-corner} shows posteriors on the parameters governing the BBH effective spin and mass ratio distributions under the base model discussed in Sect.~\ref{sec:residuals}, \textit{without} the possibility of $q-\chieff$ correlations.
Under this base model, we infer an effective spin distribution distribution centered near $\chieff \approx 0.05$ with a standard deviation of $\sigma_\chi \approx 0.1$.
Our posteriors on parameters governing the primary mass and redshift distributions are effectively identical to those presented in \citet{O3a_pop}, and so we do not show their complete posteriors here, although we include their one dimensional marginal distributions in Fig.~\ref{fig:other-params} below.

In Fig.~\ref{fig:corner-evolution} in the main text, we showed the joint posterior on parameters governing the BBH mass ratio and spin distribution under our expanded model allow for correlations between these parameters.
Figure~\ref{fig:other-params} shows the posterior on the remaining parameters when including the effects of $q-\chieff$ correlation.
There are no significant correlations between the parameters plotted in Fig.~\ref{fig:corner-evolution} and those plotted here.
Allowing for a $q-\chieff$ correlation minimally impacts our conclusions regarding the primary mass and redshift distributions of BBHs.
For comparison, the unfilled grey distributions in Fig.~\ref{fig:other-params} show posteriors obtained using the base model from Sect.~\ref{sec:residuals}; these are generally identical to the posteriors obtained from the expanded model (in blue).
The largest shift occurs in the posterior for the width $\sigma_m$ of the Gaussian peak in $m_1$.
However, $\sigma_m$ remains effectively unconstrained in both models.

\section{Recovery of Simulated Events}
\label{sec:inj-appendix}

In Sect.~\ref{sec:tests} we described an end-to-end injection study, entailing the parameter estimation and hierarchical analysis of mock signals.
This test allowed us to verify that realistic measurement degeneracies between $q$ and $\chieff$ do not seem to generically bias our conclusions regarding underlying correlations between these two parameters.
Here, we describe some additional details regarding our generation and analysis of these mock events.

We first randomly draw a large number of BBHs from a mock population with masses between $m_\mathrm{min} = 5\,M_\odot$ and $m_\mathrm{max} = 100\,M_\odot$, redshift evolution described by $\kappa  = 2$, and a mass ratio distribution with $\gamma = 0.5$.
We assume a broad $\chieff$ distribution with $\mu_{\chi,0} = 0$, $\sigma_{\chi,0} = 1$, $\alpha = 0$, and $\beta = 0$; note that this is \textit{not} the spin distribution that we ultimately inject and analyze.
Since we are concerned only with $\chieff$ and not the individual component spins, we set aligned spin components to $\chi_{1,z} = \chi_{2,z} = \chieff$ and in-plane components to zero.
We draw primary masses from a broken power law, with $p(m_1) \propto m_1^{-2}$ for $m_1<35\,M_\odot$ and $p(m_1) \propto m_1^{-4}$ for $m_1\geq35\,M_\odot$, introducing a deliberate mismatch between the mass distribution of events we inject and the model we eventually assume upon recovery.
Events are given random inclinations and sky positions.
We use \textsc{PyCBC}~\citep{pycbc} to compute the expected signal-to-noise ratio $\rho$ of each event assuming a network composed of LIGO-Hanford, LIGO-Livingston, and Virgo, using the ``O3 actual'' noise power spectral densities provided in \citet{psds} and the \textsc{IMRPhenomD} aligned-spin waveform model~\citep{Husa2016,Khan2016}.
We consider events with $\rho \geq 10$ to have been ``detected,'' and proceed until we have $5\times10^4$ such events.
This set of found events now serves two purposes.
First, it provides us a large pool from which to draw a small catalog of mock detections comparable in size to GWTC-2.
Second, this pool will allow us to quantify the appropriate selection effects when performing hierarchical inference on our mock catalog, just  as the set of pipeline injections described in Appendix~\ref{sec:hierachical-appendix} allowed us to correct for selection effects when analyzing GWTC-2.

From our set of $5\times10^4$ events, we next randomly draw 50 that will comprise our mock injection catalog.
In randomly choosing events, each event is assigned a draw weight proportional to $N(\chieff|0.05,0.15)/N(\chieff|0,1)$, so that our catalog corresponds to a much narrower effective spin distribution with $\mu_{\chi,0} = 0.05$ and $\sigma_{\chi,0} = 0.15$, rather than the broad reference distribution used above to generate our initial pool.
We use \textsc{Bilby}~\citep{Ashton2019,Romero-Shaw2020} together with the \textsc{Dynesty}~\citep{Speagle2020} nested sampler to perform parameter estimation on these 50 events, adding to each signal a random noise realization consistent with the ``O3 actual'' power spectral densities noted above.
Parameter estimation is performed using the \textsc{IMRPhenomD} aligned-spin waveform model~\citep{Husa2016,Khan2016}, with priors that are uniform in detector-frame mass and Euclidean volume [Eq.~\eqref{eq:prior-m-z}] and uniform in aligned component spins, corresponding to a $\chieff$ prior given by Eq.~\eqref{eq:prior-inj-chieff}.
This is \textit{not} the same as the $\chieff$ prior implied by a uniform and \textit{isotropic} prior on component spins, the default choice used by the LIGO and Virgo Collaborations, which is much more tightly concentrated about $\chieff = 0$.
In order to allow for a self-consistent comparison between Figs.~\ref{fig:default-posteriors-a} and \ref{fig:injection-a}, the posterior samples in Fig.~\ref{fig:injection-a} have therefore been reweighted from their original aligned spin prior to the $\chieff$ prior arising from isotropic spins \citep[Eq.~(10) of][]{Callister2021}.
%TC:endignore

{\fontsize{10}{10}\selectfont
\bibliography{References}
}

\end{document}